\documentclass[twoside,twocolumn,9pt]{article}
\usepackage{extsizes}
\usepackage[super,sort&compress,comma]{natbib}
\usepackage[version=3]{mhchem}
\usepackage[left=1.5cm, right=1.5cm, top=1.785cm, bottom=2.0cm]{geometry}
\usepackage{balance}
\usepackage{widetext}
\usepackage{times,mathptmx}
\usepackage{sectsty}
\usepackage{graphicx}
\usepackage{lastpage}
\usepackage[format=plain,justification=raggedright,singlelinecheck=false,font={stretch=1.125,small,sf},labelfont=bf,labelsep=space]{caption}
\usepackage{float}
\usepackage{fancyhdr}
\usepackage{fnpos}
\usepackage[english]{babel}
\usepackage{array}
\usepackage{droidsans}
\usepackage{charter}
\usepackage[T1]{fontenc}
\usepackage[usenames,dvipsnames]{xcolor}
\usepackage{setspace}
\usepackage[compact]{titlesec}
\usepackage{latexsym,amsmath}
\usepackage{pifont}
\usepackage{epstopdf}%This line makes .eps figures into .pdf - please comment out if not required.
\usepackage{amssymb}
\usepackage{amsfonts}
\usepackage{url}
\usepackage{ulem}

\definecolor{cream}{RGB}{222,217,201}

\usepackage{color}
\usepackage{tikz}
\usetikzlibrary{shapes,patterns,arrows}
\usepackage{booktabs}
\newcommand{\onlinecite}[1]{[\hspace{-1 ex} \nocite{#1}\citenum{#1}]}

\newcommand{\be}{\begin{equation}}
\newcommand{\ee}{\end{equation}}
\newcommand{\ba}{\begin{eqnarray}}
\newcommand{\ea}{\end{eqnarray}}

\begin{document}

\pagestyle{fancy}
\thispagestyle{plain}
\fancypagestyle{plain}{

%%%HEADER%%%
\fancyhead[C]{\includegraphics[width=18.5cm]{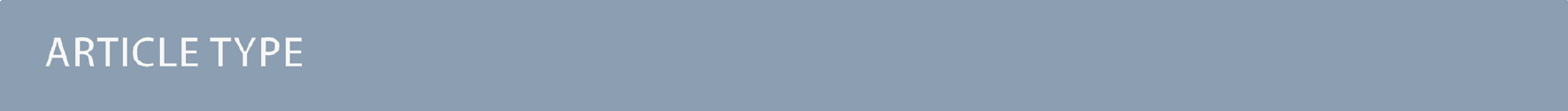}}
\fancyhead[L]{\hspace{0cm}\vspace{1.5cm}\includegraphics[height=30pt]{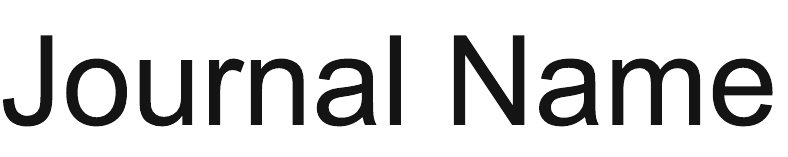}}
\fancyhead[R]{\hspace{0cm}\vspace{1.7cm}\includegraphics[height=55pt]{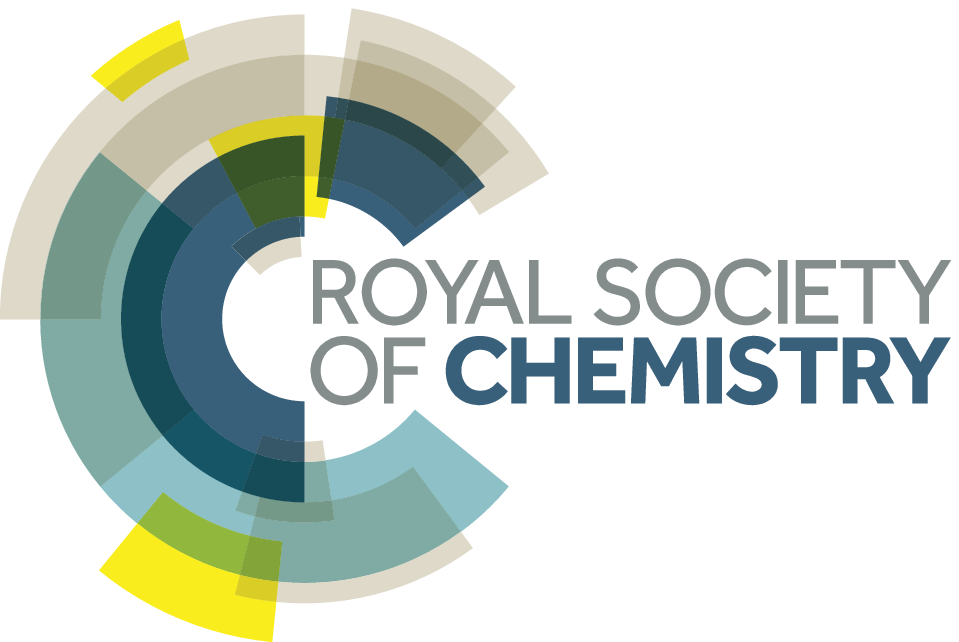}}
\renewcommand{\headrulewidth}{0pt}
}
%%%END OF HEADER%%%

%%%PAGE SETUP - Please do not change any commands within this section%%%
\makeFNbottom
\makeatletter
\renewcommand\LARGE{\@setfontsize\LARGE{15pt}{17}}
\renewcommand\Large{\@setfontsize\Large{12pt}{14}}
\renewcommand\large{\@setfontsize\large{10pt}{12}}
\renewcommand\footnotesize{\@setfontsize\footnotesize{7pt}{10}}
\makeatother

\renewcommand{\thefootnote}{\fnsymbol{footnote}}
\renewcommand\footnoterule{\vspace*{1pt}% 
\color{cream}\hrule width 3.5in height 0.4pt \color{black}\vspace*{5pt}}
\setcounter{secnumdepth}{5}

\makeatletter
\renewcommand\@biblabel[1]{#1}
\renewcommand\@makefntext[1]% 
{\noindent\makebox[0pt][r]{\@thefnmark\,}#1}
\makeatother
\renewcommand{\figurename}{\small{Fig.}~}
\sectionfont{\sffamily\Large}
\subsectionfont{\normalsize}
\subsubsectionfont{\bf}
\setstretch{1.125} %In particular, please do not alter this line.
\setlength{\skip\footins}{0.8cm}
\setlength{\footnotesep}{0.25cm}
\setlength{\jot}{10pt}
\titlespacing*{\section}{0pt}{4pt}{4pt}
\titlespacing*{\subsection}{0pt}{15pt}{1pt}
%%%END OF PAGE SETUP%%%

%%%FOOTER%%%
\fancyfoot{}
\fancyfoot[LO,RE]{\vspace{-7.1pt}\includegraphics[height=9pt]{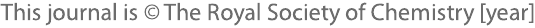}}
\fancyfoot[CO]{\vspace{-7.1pt}\hspace{13.2cm}\includegraphics{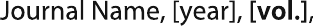}}
\fancyfoot[CE]{\vspace{-7.2pt}\hspace{-14.2cm}\includegraphics{head_foot/RF}}
\fancyfoot[RO]{\footnotesize{\sffamily{1--\pageref{LastPage} ~\textbar  \hspace{2pt}\thepage}}}
\fancyfoot[LE]{\footnotesize{\sffamily{\thepage~\textbar\hspace{3.45cm} 1--\pageref{LastPage}}}}
\fancyhead{}
\renewcommand{\headrulewidth}{0pt}
\renewcommand{\footrulewidth}{0pt}
\setlength{\arrayrulewidth}{1pt}
\setlength{\columnsep}{6.5mm}
\setlength\bibsep{1pt}
%%%END OF FOOTER%%%

%%%FIGURE SETUP - please do not change any commands within this section%%%
\makeatletter
\newlength{\figrulesep}
\setlength{\figrulesep}{0.5\textfloatsep}

\newcommand{\topfigrule}{\vspace*{-1pt}% 
\noindent{\color{cream}\rule[-\figrulesep]{\columnwidth}{1.5pt}} }

\newcommand{\botfigrule}{\vspace*{-2pt}% 
\noindent{\color{cream}\rule[\figrulesep]{\columnwidth}{1.5pt}} }

\newcommand{\dblfigrule}{\vspace*{-1pt}% 
\noindent{\color{cream}\rule[-\figrulesep]{\textwidth}{1.5pt}} }

\makeatother
%%%END OF FIGURE SETUP%%%

%%%TITLE, AUTHORS AND ABSTRACT%%%
\twocolumn[
  \begin{@twocolumnfalse}
\vspace{3cm}
\sffamily
\begin{tabular}{m{4.5cm} p{13.5cm} }

\includegraphics{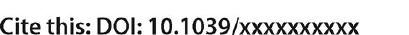} & \noindent\LARGE{\textbf{Early stages of aggregation 
in fluid
mixtures of dimers and spheres: a theoretical and simulation study}} \\%Article title goes here instead of the text "This is the title"
\vspace{0.3cm} & \vspace{0.3cm} \\

& \noindent\large{Gianmarco Muna\`o\textit{$^{1,*}$}, Santi Prestipino}\textit{$^{1}$}, and Dino Costa\textit{$^{1}$}, \\%Author names go here instead of "Full name", etc.

\includegraphics{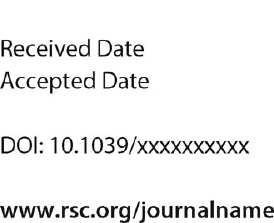} & \noindent\normalsize{
We use Monte Carlo simulation and the Reference Interaction Site Model (RISM)
theory of molecular fluids to investigate a simple model of colloidal mixture
consisting of dimers, made up of two tangent hard monomers of different size,
and hard spheres. In addition to steric repulsion, the two
species interact via a square-well attraction only between small 
monomers
and spheres. Recently, we have characterized 
the low-temperature regime of this mixture
by Monte Carlo, reporting on the spontaneous formation of a
wide spectrum of supramolecular aggregates 
[Prestipino et al, J. Phys.  Chem. B, 2019, {\bf 123}, 9272]. 
Here we focus on a regime of temperatures
where, on cooling, the appearance of local inhomogeneties first, and the
early stages of aggregation thereafter, are observed.
In particular, we find signatures
of aggregation in the onset of a low-wavevector peak in the structure
factors of the mixture, as computed by both theory and simulation. Then, we link
the structural information to the microscopic arrangement through a 
detailed cluster analysis
of Monte Carlo configurations. In this regard, we devise a novel method 
to compute the maximum distance for which two spheres
can be regarded as bonded together, a crucial issue in the
proper identification of fluid aggregates. The RISM theory provides
relatively accurate structural and thermodynamic predictions in comparison
with Monte Carlo, but with slightly degrading performances as the fluid progresses
inside the locally inhomogeneous phase. Our study certifies the efficacy
of the RISM approach 
as a useful complement to numerical simulation for a reasoned analysis
of aggregation properties in colloidal mixtures.}

\end{tabular}

 \end{@twocolumnfalse} \vspace{0.6cm}

  ]
%%%END OF TITLE, AUTHORS AND ABSTRACT%%%

%%%FONT SETUP - please do not change any commands within this section
\renewcommand*\rmdefault{bch}\normalfont\upshape
\rmfamily
\section*{}
\vspace{-1cm}

%%%FOOTNOTES%%%

\footnotetext{\textit{$^{1}$~Dipartimento di Scienze Matematiche e Informatiche, Scienze Fisiche e Scienze della Terra, Universit\`a degli Studi di Messina, Viale F.~Stagno d'Alcontres 31, 98166 Messina, Italy.}}

\footnotetext{\textit{$^{*}$~Corresponding author, e-mail: gmunao@unime.it}}

%\footnotetext{\textit{$^{2}$~CNR-IPCF, Viale F. Stagno d'Alcontres 37, 98158 Messina, Italy.}}

%%%END OF FOOTNOTES%%

\section{Introduction}

At the mesoscale, complex fluids can be characterized as systems capable to self-assemble, i.e. 
to spontaneously form organized structures out of simpler building blocks. 
The comprehensive study of these patterns, along with the possibility to control their shape 
and size by tuning the interaction law, is among the greatest challenges of soft matter physics. 

Recent studies have shown that a rich variety of mesoscopic patterns are obtained for
asymmetrically-shaped particles whose constituent blocks interact 
through spherically-symmetric potentials.~\cite{Avvisati:15,Avvisati-Soft,Mittal:15,Mittal:16} 
In this regard, a versatile model is represented by Janus 
dimers, i.e. heteronuclear dimers in which
one monomer is solvophilic and the other one is solvophobic.~\cite{Janusdumbbell,Janusdumbbell1} 
The interest in this specific class of colloidal particles stems from
their capability to be synthesized in a large variety of
sizes, aspect ratios, and interaction 
properties,~\cite{Kim:06,Nagao:10,Chakrabortty:10,Skelhon:14,Yoon-Chem} and
from their high
technological impact.~\cite{Forster,Hosein,Ma:12} 
Moreover, Janus dimers are able to self-organize 
into various kinds of supracolloidal structures, provided that
appropriate thermodynamic conditions are met,~\cite{Kraft:12,Bon:14,Hu:16} as also observed in 
theoretical and simulation studies.~\cite{Munao:14,Barbosa:15,Munao:JPCM}

Recently, we have undertaken an extended investigation of the phase behavior of a dilute colloidal 
mixture of Janus dimers 
and spherical particles,
by means of computer simulation.
%~\cite{Munao:capsule,Prestipino1,Prestipino:19,Prestipino2} 
~\cite{Munao:capsule,Prestipino:17b,Munao:17,Prestipino:19} 
In our scheme, a dimer is modeled
as a pair of tangent hard spheres of different size, while the other species
is represented by a hard sphere. Beside the steric repulsion, the two species
interact via a square-well attraction between small monomers and spheres.
We take the square-well width to be 
small compared to the size of spheres, as is usual for colloidal systems, 
but at the 
same time large enough to allow for the formation of aggregates of 
dimers and spheres at low temperature.~\cite{Prestipino:19} 
Moreover, the absence of any dimer-dimer or sphere-sphere attraction
reflects the implicit assumption that this kind of interaction is significantly
weaker that the dimer-sphere attraction.  
Finally,  
the significant size asymmetry
between the monomers constituting the
dimers favors the encapsulation of spheres.~\cite{Prestipino:17b}
Indeed, we have seen that,
for sufficiently low densities and temperatures,
spheres gather together via an effective mutual attraction mediated
by small monomers, whereas the uncontrolled growth of aggregates~---~%
followed by coarsening into a macroscopic droplet~---~%
is prevented by the hindrance exerted by large monomers, which form a protective coating
around the hard-sphere aggregates.
As a result, liquid-vapor separation is preempted by the formation
of supramolecular aggregates.
We have documented a rich phase scenario arising for our mixture,
in terms of diameter and concentration of spheres:~\cite{Prestipino:19}
when the two species have similar sizes and the concentration is small,
we observe the onset of small clusters of spheres covered by a 
layer of dimers; for larger concentrations, we find other self-assembled structures 
(i.e. gel-like networks and bilayers); finally, when spheres are
substantially larger than dimers, we observe the formation of membranes (i.e. curved sheets)
and occasionally vesicles. A wide variety of aggregates also exists in two dimensions,
both on a plane~\cite{Prestipino:17} and on a spherical surface.~\cite{Dlamini:21}

Recently, a colloidal mixture with characteristics similar to ours
has been synthesized and investigated experimentally.~\cite{Wolters:17} At variance with
the simulated mixture, real spheres tend
to coalesce, whereas the inclusion of Janus dimers 
stabilizes them against aggregation, with an encapsulation mechanism
in gratifying accordance 
with our simulation results.

In this paper we study the fluid phase behavior of our
mixture, at fixed total density (higher than in previous studies)
and as a function of the sphere concentration.
Upon lowering the temperature from high values, 
local inhomogeneities first appear, thereafter followed by the formation of aggregates.
Owing to the large number of parameters involved
and due to long relaxation times, a systematic analysis of the fluid sector would greatly benefit from
the availability of a theoretical scheme complementing the much more costly simulation approach. 
More generally, 
having reliable theoretical tools to perform a quick analysis of the 
structure and thermodynamics of a complex fluid with many parameters is of uppermost importance.

With this in mind, we assess the performance of the Reference Interaction Site Model
(RISM) theory of molecular fluids,~\cite{chandler:1972}
which is benchmarked against the numerical data generated by Monte Carlo. 
The RISM formalism is one of the most effective tools to investigate the structure and thermodynamics of 
molecular fluids. Derived as a molecular generalization of the Ornstein-Zernike theory of simple fluids,~\cite{Hansennew} 
the RISM approach was originally intended for rigid molecules made up of hard spheres.~\cite{lowden:5228} 
Later on, the theory was extended to more general systems by the inclusion of long-range forces and 
attractive interactions. With these improvements, the RISM theory was successfully applied 
for associating fluids like water~\cite{Lue:5427,Kovalenko:01} and other polar
solvents.~\cite{pettitt:7296,costa:224501} Recently, RISM has been employed to investigate
the statistical properties of colloids,~\cite{Harnau:01,Harnau:05,Munao:11}
polymers,~\cite{Hansen:06}, macromolecules,~\cite{Khalatur:97} and nanoparticles,~\cite{Kung:10}
We have used the RISM theory to study pure fluids of
dimers, finding a reasonable agreement
with simulation results and other theoretical approaches.~\cite{Munao:PCCP,Munao:15}

As argued before, the origin of inhomogeneities in the sphere subsystem
can be traced back to a competition between a (dimer-mediated) short-range attraction
and a longer-range (again effective) repulsion. The same mechanism is actually at work in
one-component fluids of SALR (Short-range Attractive and Long-range Repulsive) particles.
While in SALR fluids the competing interactions are directly encoded in the shape
of the isotropic pair potential, in our system the interplay between attraction and repulsion
arises from the interaction between the species.
SALR fluids are currently employed to mimic the behavior of a variety of 
soft materials~\cite{Baglioni:04,Campbell:05,Cardinaux:07,Schurt:04,Godfrin:15,Riest:15,Riest:18}
and, as such, they are largely investigated in the current 
literature; four recent reviews witness the broad interest 
in this topic.~\cite{Zhuang:16,Sweatman:19,Liu:19,Bretonnet:19}
In view of the intrinsic similarity of our mixture with a SALR fluid, we will analyze
the former system by the same tools applied for the latter. In particular, since the onset
of aggregation in a SALR fluid is generally manifested in the appearance of a
low-wavevector peak in the static structure factor,~\cite{liu:11,falus:12,Godfrin:13,Godfrin:14}
we expect a similar signature to occur in the various (partial and total)
structure factors of the mixture. Among them, we will analyze those which
convey the most sensitive information about the microscopic state of the mixture.
The interpretation of structural data in terms of the characteristics of particle arrangement
is another well established step in the study of SALR fluids. Here, we follow
a similar approach for the mixture, by relying on the cluster analysis of Monte Carlo configurations
as a means to map out the mesoscale structures.
In this regard, we devise a simple and robust criterion to determine  
the maximum distance for which two spheres can be considered
as bonded together, i.e. belonging to the same aggregate.
This parameter is crucial in the attempt to uncover
the nature of aggregates developing in the fluid.

The outline of the paper is the following. After describing the model and 
methods 
in Section~2, we present and discuss our results in Section~3. Concluding remarks 
and perspectives for future studies are reported in Section~4.
\section{Model and methods}

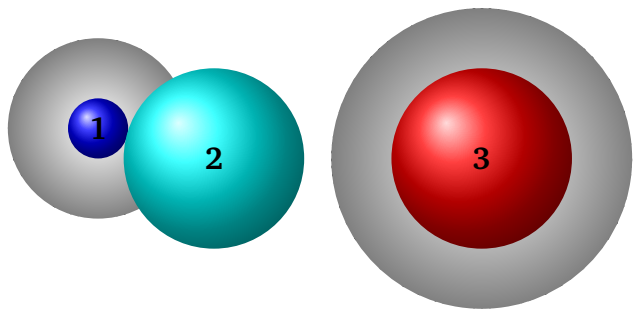
\begin{figure}
\begin{tikzpicture}
% DIMER, SITE 1
% halo
\node[pattern=dots,circle,outer color=gray,inner color=white,minimum width=2.4cm] (radial) at (-3.4,2.4) {};
% particle
\node[circle,shading=ball,ball color=blue,minimum width=0.8cm] (ball) at (-3.4,2.4) {\Large{\bf 1}};
% DIMER, SITE 2
\node[circle,shading=ball,ball color=cyan,minimum width=2.4cm] (ball) at (-1.86,2.0) {\Large{\bf 2}};
% HS
% halo
\node[pattern=dots,circle,outer color=gray,inner color=white,minimum width=4.0cm] (radial) at (1.7,2.0) {};
% particle
\node[circle,shading=ball,ball color=red,minimum width=2.4cm] (ball) at (1.7,2.0) {\Large{\bf 3}};
\end{tikzpicture}
\caption{The species making up our mixture: A dimer consists of a small (1, blue) and a large monomer (2, cyan) 
in a size ratio of $\sigma_1/\sigma_2=1/3$; the spherical particle (3, red) has diameter $\sigma_3=\sigma_2$. 
The grey halos around particles 1 and 3 represent the attractive region, 
extending up to $\sigma_1$ beyond the distance of closest approach $\sigma_{13}$, see Eq.~(\ref{eq:pot}).}
\label{fig:models}
\end{figure}

In our system, a dimer is modeled as two tangent hard spheres
with different diameters, $\sigma_1$ and $\sigma_2$, in a fixed ratio 
$\sigma_2=3\sigma_1$. Spherical particles are represented as 
hard spheres of diameter $\sigma_3=\sigma_2$, see Fig.~\ref{fig:models}.
All interactions are hard-sphere-like with additive diameters
$\sigma_{ij}=(\sigma_i+\sigma_j)/2$, except for the interaction between
the small monomer and the sphere, which is given the form of a square-well
attractive potential:
\ba
u_{13}(r)=
\begin{cases}
\infty & \text {if} \ r<\sigma_{13} \\[4pt]
-\varepsilon & \text{if}\  \sigma_{13}\le r<\sigma_{13}+\sigma_1\\[4pt]
0 & \text {otherwise}
\end{cases}
\label{eq:pot}
\ea
In the following, 
$\sigma_2$ and $\varepsilon$ are taken as units of length and energy, 
respectively, which in turn leads to a reduced number density $\rho^\ast=\rho\sigma_2^3$
and a reduced temperature $T^\ast=k_{\rm B}T/\varepsilon$, $k_{\rm B}$ being Boltzmann's constant. 
Finally, we denote $N_{\rm d}$ and $N_{\rm s}$ the number of dimers and spheres, respectively. 
Hence, $N=N_{\rm d}+N_{\rm s}$ is the total number of particles, $\chi_{\rm d}=N_{\rm d}/N$
and $\chi_{\rm s}=1-\chi_{\rm d}$ are the relative concentrations of the two species,
and we set $\chi_{\rm ds}=\sqrt{\chi_{\rm d}\chi_{\rm s}}$.

%%%%%%%%%%%%%%%%%%%%%%%%%%%%%%%%%%%%%%%%%%%
%
%   FIGURA 1
%
%%%%%%%%%%%%%%%%%%%%%%%%%%%%%%%%%%%%%%%%%%%

To explore the behavior of the model in the fluid phase we employ
the RISM approach: given an arbitrary molecular species, represented as a
geometric assembly of $n$ distinct interaction sites (like in a ball-and-stick model),
the RISM theory provides a relationship between the set of 
$n(n+1)/2$ site-site total correlation functions
$h_{ij}(r)=g_{ij}(r)-1$~---~with $g_{ij}(r)$ the radial distribution functions relative
to sites $i$ and $j$ of different molecules~---~%
and the set of direct correlation functions, $c_{ij}(r)$.
In reciprocal (wavevector) space, the RISM equation reads:~\cite{Hansennew}
\be
{\bf H}(k)={\bf W}(k){\bf C}(k){\bf W}(k)+\rho{\bf W}(k){\bf C}(k){\bf H}(k)\,,
\label{eq:rism}
\ee
where ${\mathbf H}\equiv [h_{ij}(k)]$ and  ${\mathbf C}\equiv [c_{ij}(k)]$ 
are $n\times n$ symmetric matrices and 
$\rho$ is the number density.
The RISM formalism can be derived as a generalization of the Ornstein-Zernike relation for a
mixture of monatomic species, where a matrix of intramolecular correlation
${\mathbf W}\equiv[w_{ij}(k)]$ is introduced to account for the bonds
between the interaction sites of a molecule. In particular, 
\ba
w_{ij}(k) = \dfrac{\sin(kL_{ij})}{kL_{ij}}\,,
\label{eq:w}
\ea
where $L_{ij}$ is the bond distance between sites $i$ and $j$ of the same molecule.
The RISM equation~(\ref{eq:rism}) must be complemented with a ``closure'' relation,
to form a closed set of equations; we have adopted to this scope
the hypernetted-chain (HNC) approximation,~\cite{Hansennew}
\be
c_{ij}(r)=\exp[-\beta u_{ij}(r)+\gamma_{ij}(r)]-\gamma_{ij}(r)-1\,,
\label{eq:hnc}
\ee
where $\gamma_{ij}(r)=h_{ij}(r)-c_{ij}(r)$ and $\beta=1/T^*$.
HNC shows good performances
when applied to the study of colloids~\cite{Munao:PCCP}
and cluster-forming liquids:~\cite{Bomont:20a};
nevertheless, in the course of our study we have assessed 
its accuracy in comparison with other common
closures, such as the Percus-Yevick and 
the Mean-Spherical Approximation.~\cite{Hansennew} 
In this way, 
we have ascertained the clear superiority of
HNC for the mixture at issue, especially in the presence of aggregation, 
where the other schemes tend to drastically overlook 
the progressive structuring of the fluid.

Equation~(\ref{eq:rism}) can be generalized to mixtures of molecular species
and then specialized to the dimer-sphere system under study. In our case,
all the matrices entering the formalism are $3 \times 3$ matrices, and
the intramolecular correlation matrix is explicitly given by
\be
{\bf W}(k)=
\begin{bmatrix}
	1 & \dfrac{\sin(k\sigma_{12})}{k\sigma_{12}} & 0 \\[4pt]
	\dfrac{\sin(k\sigma_{12})}{k\sigma_{12}}  & 1 & 0 \\[8pt]
0 & 0 & 1
\end{bmatrix}
\ee

Finally, the density $\rho$ in Eq.~(\ref{eq:rism}) is replaced by a diagonal matrix
$D$ with elements $D_{11}=D_{22}=\rho\chi_{\rm d}$ and
$D_{\rm 33}=\rho\chi_{\rm s}$.

We solve the set of coupled RISM/HNC equations 
numerically by means of an iterative Picard algorithm, using a real-space grid of 
$2^{14}$ points and a mesh step of $\Delta r=0.001\sigma_2$. 

\begin{figure*}
\begin{center}
\begin{tabular}{cc}
\includegraphics[width=6.8cm,angle=-90]{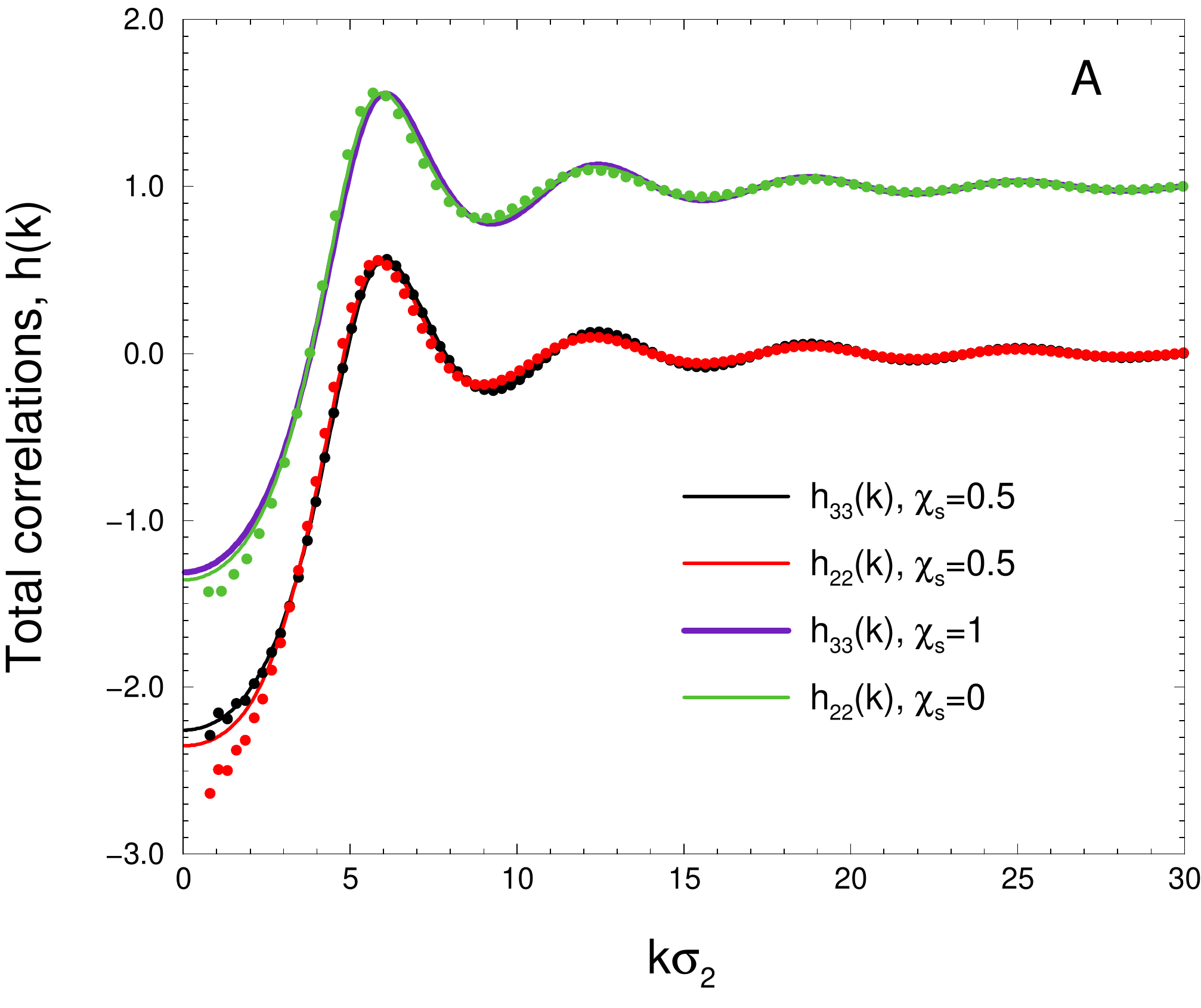}  &
\includegraphics[width=6.8cm,angle=-90]{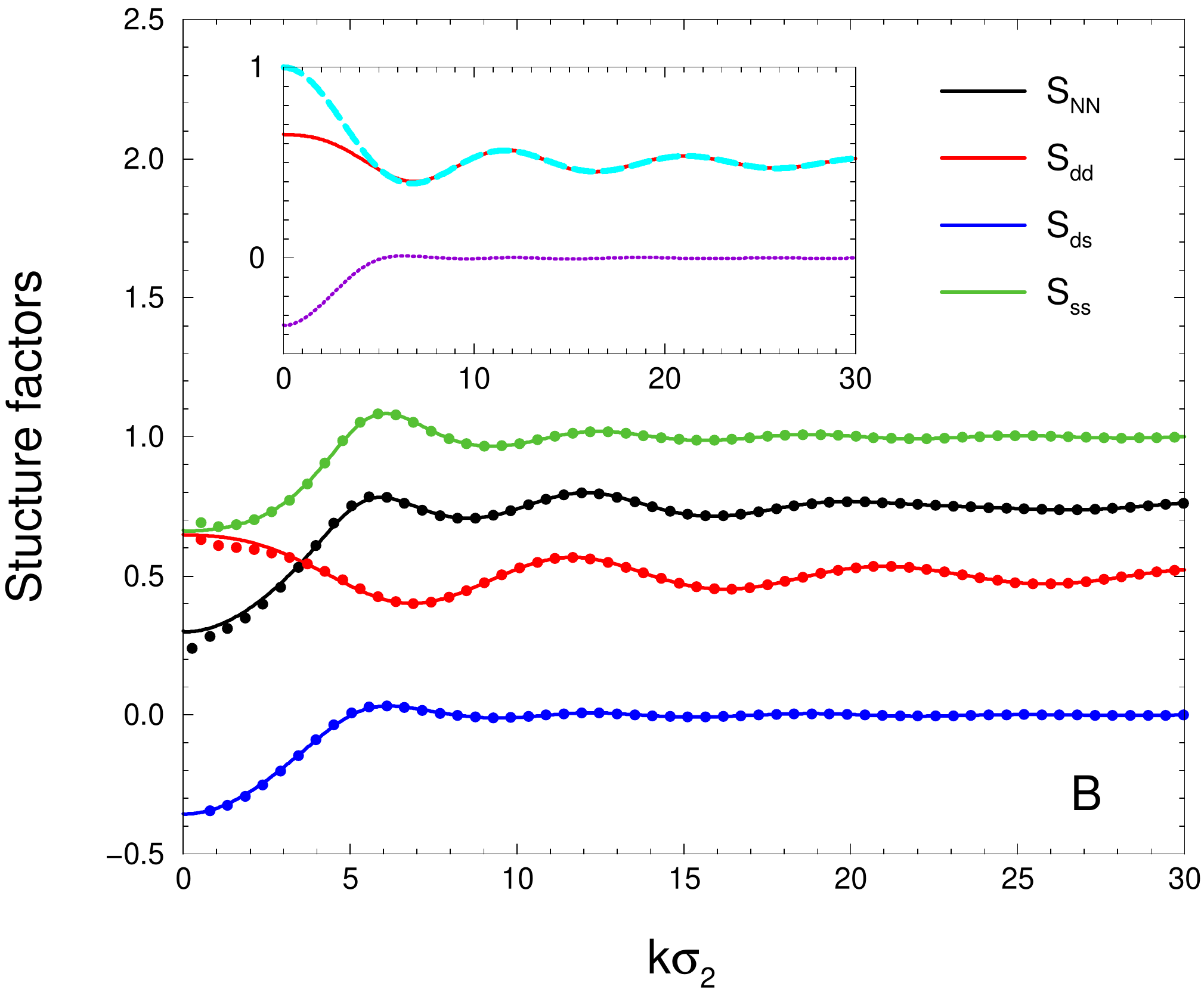} 
\end{tabular}
\caption{Mixture of dimers and spheres at infinite temperature and $\rho^\ast=0.30$.
A: RISM predictions (lines) and MC data (symbols)
for ${h}_{33}(k)$ and ${h}_{22}(k)$ ($\chi_s$ values are given
in the legend); for clarity, the curves for pure fluids are vertically shifted by one.
B: RISM (lines) and MC (symbols) for the various structure factors defined
in Eqs~(\ref{eq:sdd})-(\ref{eq:snn}), for $\chi_{\rm s}=0.5$.
In the inset, $S_{\rm dd}(k)$ (full red line) has been resolved into
intramolecular [$F(k)$, dashed cyan line] and intermolecular [$S_{\rm dd}(k)$, 
dotted indigo line]
contributions, see Eq.~(\ref{eq:sdd}).
}
\label{fig:highT}
\end{center}
\end{figure*}
We shall present most of our structural results in terms of $k$-space correlations,
i.e. in terms of the various structure factors characterizing the mixture. 
The dimer-dimer structure factor $S_{\rm dd}(k)$ reads:
\ba\label{eq:sdd}
S_{\rm dd}(k) &=&
\dfrac{1}{2^2} \sum_{i,j=1,2} w_{ij}(k) + \dfrac{\rho\chi_{\rm d}}{2^2}\sum_{i,j=1,2}h_{ij}(k) \nonumber \\
&\equiv & F(k)+S'_{\rm dd}(k)\,.
\ea
The first term on the r.h.s. of (\ref{eq:sdd}) is the molecular form factor $F(k)$,
which characterizes the shape of the dimer.
The form factor accounts for the interference
of radiation scattered from different parts of the same
particle in a diffraction experiment.
The correlations between different dimers 
are instead expressed by $S'_{\rm dd}(k)$, which depends
on the Fourier transform of the total correlation functions
relative to monomers only.
The sphere-sphere structure factor is
\be\label{eq:sss}
S_{\rm ss}(k)\equiv S_{33}(k)=1+\rho\chi_{\rm s}{h}_{33}(k)\,,
\ee
and the dimer-sphere structure factor, involving the Fourier transform
of the cross-correlation functions $h_{13}(r)$ and $h_{23}(r)$, is given by
\be\label{eq:sds}
S_{\rm ds}(k)=\dfrac{1}{2} \rho \chi_{\rm ds} \ [{h}_{13}(k)+{h}_{23}(k)]\,.
\ee 

In terms of the partial structure factors $S_{\rm dd}(k)$, $S_{\rm ss}(k)$,
and $S_{\rm ds}(k)$, we can also define ``total'' structure factors,
describing correlations between fluctuations of global variables,
like for instance the total number density:
\be\label{eq:snn}
 S_{\rm NN}(k)  = \chi_{\rm d} S_{\rm dd}(k) +
2\chi_{\rm ds} S_{\rm ds}(k)+ \chi_{\rm s} S_{\rm ss}(k)\,,
\ee
which is hereafter referred to as the total structure factor 
(see Ref.~\onlinecite{March} for details).

Moving to thermodynamic quantities, the isothermal
compressibility $K_{\rm T}$ is given as a combination
of $k \to 0$ limits of the partial structure factors:~\cite{Ashcroft:67,Hansennew}
\be\label{eq:comp}
\rho k_{\rm B}^{\ }T K_{\rm T}= \dfrac
{ S_{\rm dd}(0)S_{\rm ss}(0)-[S_{\rm ds}(0)]^2}
{\chi_{\rm s} S_{\rm dd}(0)+\chi_{\rm d} S_{\rm ss}(0)
- 2\chi_{\rm ds} S_{\rm ds}(0)}\,.
\ee
Finally, for a generic interaction-site model the internal energy reads
\be\label{eq:uex}
\dfrac{U}{N}=2 \pi \rho \sum_{i,j} \chi_i \chi_j \int_0^{\infty}
u_{ij}(r) g_{ij}(r) r^2 dr \,,
\ee
where the sum runs over all interaction sites. For the present mixture,
this formula is simply reduced to
\be\label{eq:uexred}
\dfrac{U}{N\varepsilon}=-4\pi\rho\chi_{\rm d}\chi_{\rm s}
\int_{\sigma_{13}}^{\sigma_{13}+\sigma_1} g_{13}(r) r^2 dr \,.
\ee

In the present work, RISM predictions are assessed against
canonical-ensemble Monte Carlo (MC) simulations of a sample
of $N=1372$ particles enclosed in a cubic box with periodic boundary conditions.
A larger sample of 4000 particles is occasionally considered to estimate
the size dependence of the results. As a general protocol,
we have employed a million MC cycles to equilibrate
the system, followed by twice longer productions runs.
In our scheme, a MC cycle involves $N$ trial single-particle moves;
for a dimer, either a trial displacement or a trial re-orientation is randomly attempted.
The orientational move is implemented by the Barker and Watts
procedure,~\cite{Barker:69,AllenTildesley} consisting in a trial rotation around a randomly chosen
coordinate axis. The maximum extent of a displacement and the maximum
rotation angle are adjusted during the equilibration stage in such a way
as to ensure an acceptance ratio between 40\% and 60\%.

With specific concern to the hard-sphere subsystem,
we have computed several properties to characterize the presence of
local inhomogeneities and aggregates. 
In order to identify connected assemblies of hard spheres,
we have used the Hoshen-Kopelman algorithm,~\cite{Hoshen:76}
readily generalized to work with an off-lattice system of 
particles. A thorough discussion of the optimal choice of the
``bond distance'' between two hard spheres is deferred to the
next Section. The cluster-size distribution, $N(s)$, is defined as in 
Ref.~\onlinecite{Chen:94,Godfrin:14}:
\be\label{eq:csd}
N(s)=\frac{s}{N_{\rm s}}\langle n(s)\rangle \,,
\ee
where $\langle n(s) \rangle$ is the average number of clusters
with size $s$
per single configuration;
the normalization of $N(s)$ is such that $\sum_{s} N(s)=1$.
The cluster analysis is typically carried out on a set
of 1000 configurations taken from the last part of the production run.
We also monitor the total number of clusters, the fraction of isolated particles,
the size of the largest aggregate, and the number of bonds
per particle.

\begin{figure*}
\begin{center}
\begin{tabular}{cc}
\includegraphics[width=6.8cm,angle=-90]{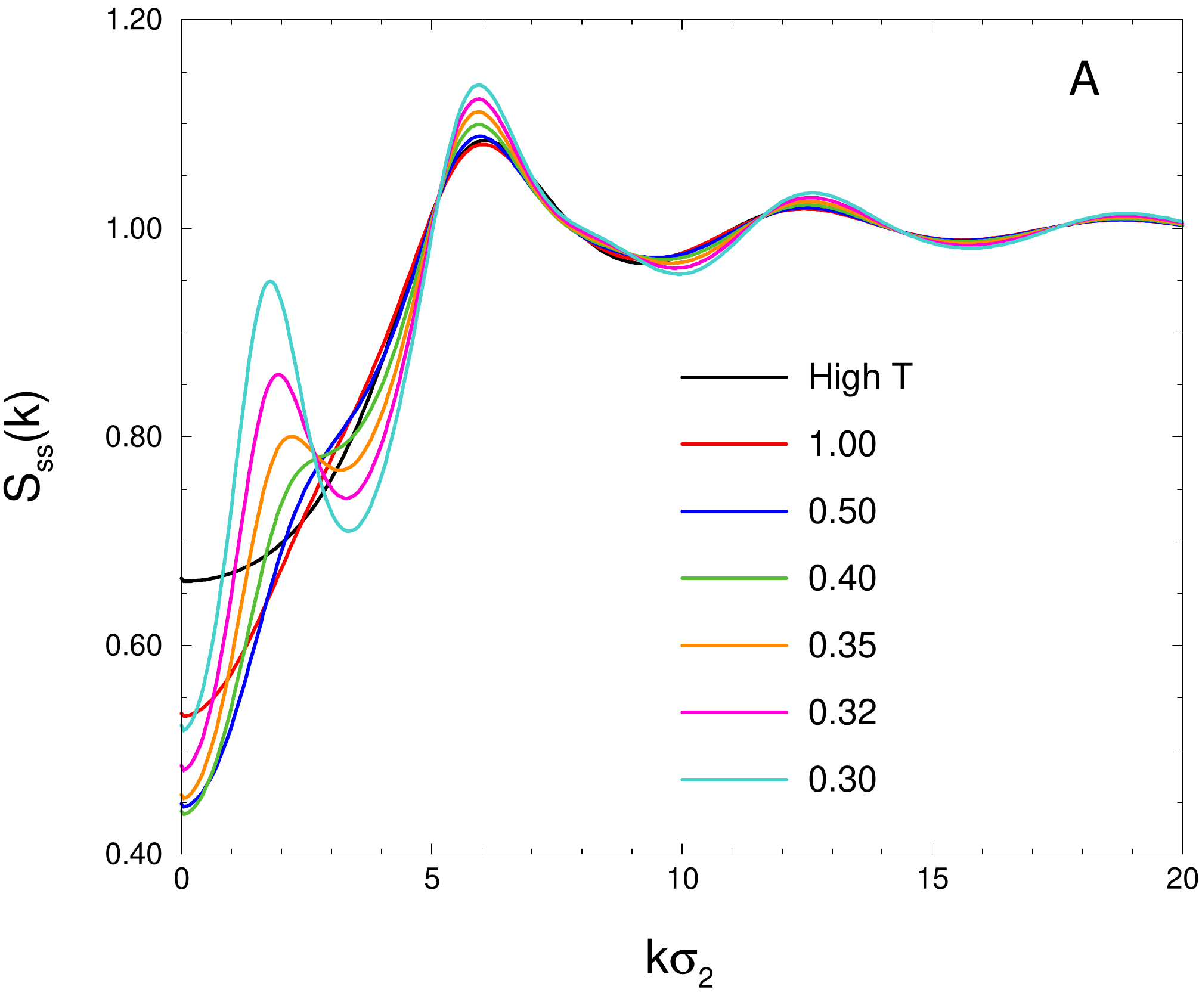} &
\includegraphics[width=6.8cm,angle=-90]{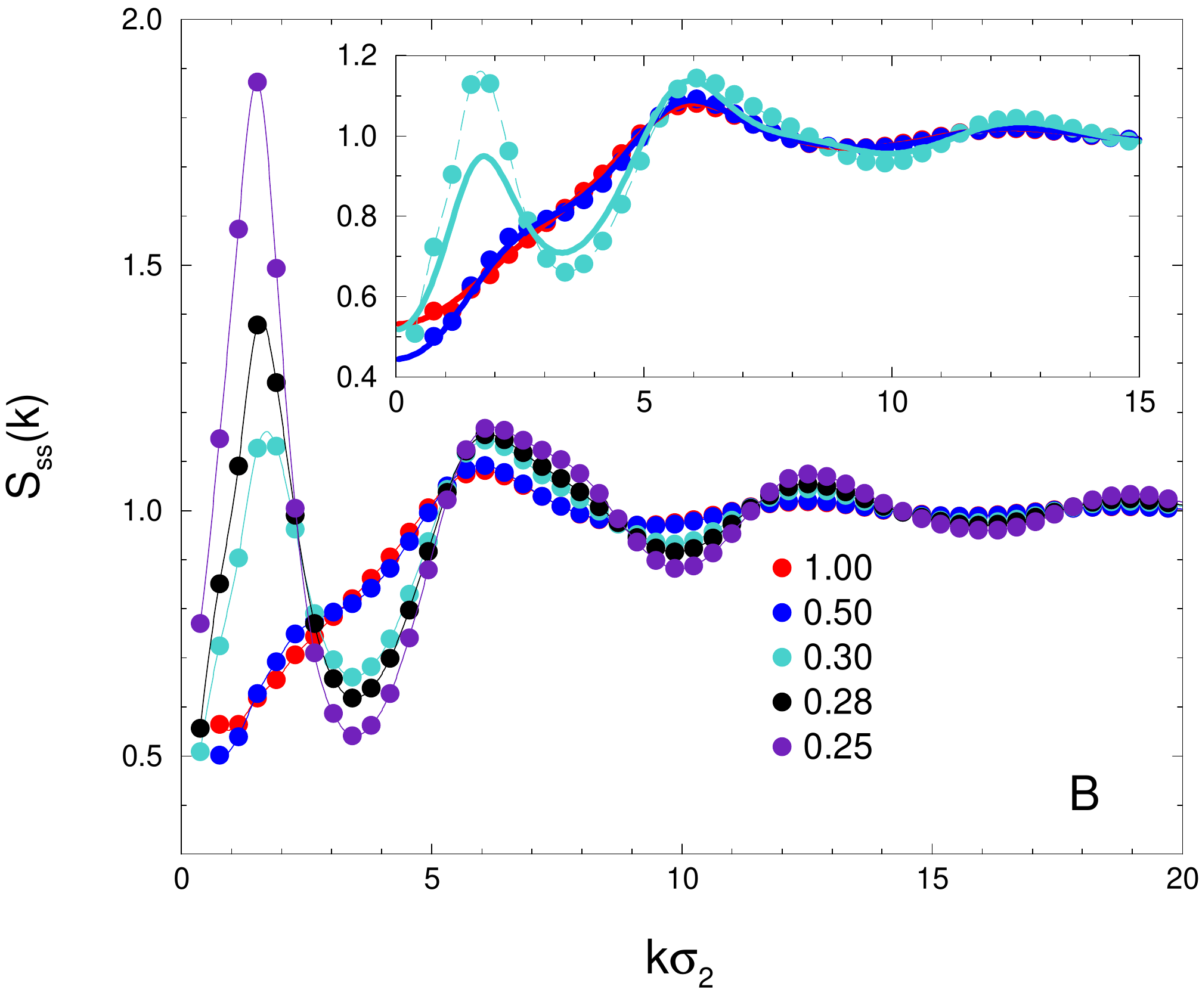} 
\end{tabular}
\caption{Sphere-sphere structure factor $S_{\rm ss}(k)$
at equimolarity and progressively lower temperatures (see the legends).
RISM (A) and MC (B, symbols, with lines as guides to the eye) are shown.
Inset: comparison between RISM predictions (thick lines) and MC data (symbols,
with dashed lines as guides to the eye).}
\label{fig:lowT}
\end{center}
\end{figure*}

\section{Results}

We start discussing the accuracy of RISM calculations for the equimolar
mixture, i.e. $\chi_{\rm s}=\chi_{\rm d}=0.5$, using MC simulation data as a reference.
Under equimolarity conditions, a quick survey indicates that RISM works best
for densities in the range $0.20-0.30$, while the theoretical predictions
become less accurate for lower densities (for $\rho^\ast\lesssim 0.1$).
Hence, from now on we will focus our analysis
on the thermodynamic states with total density $\rho^\ast=0.30$.

In order to establish a useful benchmark,
we consider a mixture of dimers and spheres at infinite temperature,
where the square-well attraction is ineffective and only steric repulsions survive.
Under these conditions, the fluid behaves as a homogeneous 
mixture of non-interacting hard dimers and hard spheres.
Total correlations between spheres only $[h_{\rm 33}(k)]$
and large monomers only $[h_{\rm 22}(k)]$ 
are shown in Fig.~\ref{fig:highT}A, as computed by RISM and MC.
Perhaps not surprisingly, the two functions are practically coincident,
even for $\chi_{\rm s}=1$ (pure spheres) and $\chi_{\rm s}=0$ (pure dimers),
due to the negligible contribution of small monomers to the
packing fraction of the mixture.
Overall, RISM predictions agree reasonably well with MC data,
except for a slight overestimate of the correlations
at small wavevectors; it is clear that these discrepancies
are due to the inescapable approximations of the theory,
as well as to the known difficulty to get accurate results
in this regime from simulation.

The profile of the total structure factor $S_{\rm NN}(k)$ is shown in
Fig.~\ref{fig:highT}B.
All the terms entering Eq.~(\ref{eq:snn}) are also reported in the figure.
It appears that the large $k$ oscillations in $S_{\rm NN}(k)$ are mainly
determined by the dimer-dimer contribution, which in turn reflects the
properties of the form factor $F(k)$, as documented in the inset. 
Looking at the definitions, the various structure factors tend to different limits
for large $k$:  $S_{\rm dd}(k)$ has the same limiting value of $F(k)$ [1/2 in our case, 
i.e. the inverse number of molecular sites, see Eq.~(\ref{eq:sdd})]; 
$S_{\rm ss}(k) \to 1$, see Eq.~(\ref{eq:sss}); $S_{\rm ds}(k) \to 0$, see Eq.~(\ref{eq:sds});
therefore, $S_{\rm NN}(k) \to \chi_{\rm d}/2+\chi_{\rm s}=0.75$ for an equimolar
mixture, see Eq.~(\ref{eq:snn}).
We can appreciate the substantial agreement between RISM and MC, at least
in the considered conditions, thus confirming the very good performance of the theory,
which, we recall, was originally designed to describe correlations in purely steric fluids. 

As the temperature is lowered, the square-well attraction becomes increasingly
relevant and, in parallel, also the tendency of hard spheres to aggregate becomes
more and more evident.
In particular, a clear signature of aggregation is, as in SALR fluids, the onset of
a low-$k$ peak in $S_{\rm ss}(k)$, i.e. a peak preceding the main diffraction peak,
see Fig.~\ref{fig:lowT}; here, RISM predictions (in panel A)
are contrasted with MC results (panel B), with a more stringent comparison
between the two in the inset. We see that the low-$k$ peak falls at $k_0\lesssim 2$,
i.e. well below the main diffraction peak at $\approx 2\pi/\sigma_{33}$.
Looking first at the RISM results, the onset of local inhomogeneities is signaled
by the development of an inflection point in the low-$k$ region for temperatures
around $T^\ast\approx 1.0$, which is then transformed into a shoulder for
$T^\ast \approx 0.50$, before finally evolving into a well-shaped peak for
$T^\ast \approx 0.35$. As the temperature is further lowered, the peak grows
in height until reaching a maximum of $\approx 0.95$ for $T^\ast=0.30$.
If we now try to reduce the temperature further, we run into a drawback,
since the RISM algorithm fails to converge to a physically meaningful solution
(typically, large oscillations show up in real-space correlations
even for large distances or deep negative peaks appear in the 
structure factors). 
This outcome is not unexpected
since RISM~---~like any other theory designed for
homogeneous fluids~---~%
only works for sufficiently high temperatures, where the spatial inhomogeneities are
not too marked (or where the fluid does not enter 
too much within a liquid-vapor phase separation).
Clearly, this problem
does not affect simulation, and the subsequent MC evolution of the
sphere-sphere structure factor below $T^\ast=0.30$ is reported in
Fig.~\ref{fig:lowT}B.
The small downward shift in the position of the low-$k$ peak on cooling,
which is observed both in the RISM and in MC data, signals
a corresponding reorganization of aggregates over larger distances.
The distribution of dimers in space follows the same trend of spheres:
in particular, a low-$k$ peak emerges on cooling also in the functions
${h}_{13}(k)$ and ${h}_{23}(k)$ (not shown).
This similarity of behavior is a clear manifestation of the crucial role of
small monomers in providing the bridging ``glue'' between spheres, hence 
the structure factors of both species have the same gross features. 
The comparison between theory and simulation, reported in the inset of
Fig.~\ref{fig:lowT}B, shows that RISM closely follows the MC evolution
of $S_{\rm ss}(k)$, even at the lowest temperature attainable by theory.
Only the height of the low-$k$ peak is slightly underestimated,
which again reflects 
the increasing difficulties encountered by RISM in keeping up with the
enhancing of inhomogeneities on cooling.

The way how incipient aggregation is reflected in the various
contributions making up the total structure factor $S_{\rm NN}(k)$
is shown in Fig.~\ref{fig:snn}, referring to RISM calculations for
$T^\ast=0.30$. Here, a low-$k$ peak is present not only in $S_{\rm ss}(k)$,
but also in the structure factors involving dimers, i.e. $S_{\rm dd}(k)$ and $S_{\rm ds}(k)$.
As a result, {$S_{\rm NN}(k)$ exhibits a local maximum at 
$k\approx 1.5$;
the latter position is an average determined from the superposition of various
structure-factor peaks located between 1 and 2. More generally, the profile of
$S_{\rm NN}(k)$ faithfully follows that of the cross contribution $S_{\rm ds}(k)$
over the whole $k$ range. This should be contrasted with the previous observation that,
for purely steric interactions, $S_{\rm NN}(k)$ closely reflects $S_{\rm dd}(k)$.
For temperatures lower than $T^\ast=0.30$, i.e. out of the reach of RISM, 
MC results (reported in the inset of Fig.~\ref{fig:snn}) show that the low-$k$
peak grows on cooling.  

%%%%%%%%%%%%%%%%%%%%%%%%%%%%%%%%%%%%%%%%%%%%%%%%%%%%%%%%
\begin{figure}
\begin{center}
\begin{tabular}{c}
\includegraphics[width=6.8cm,angle=-90]{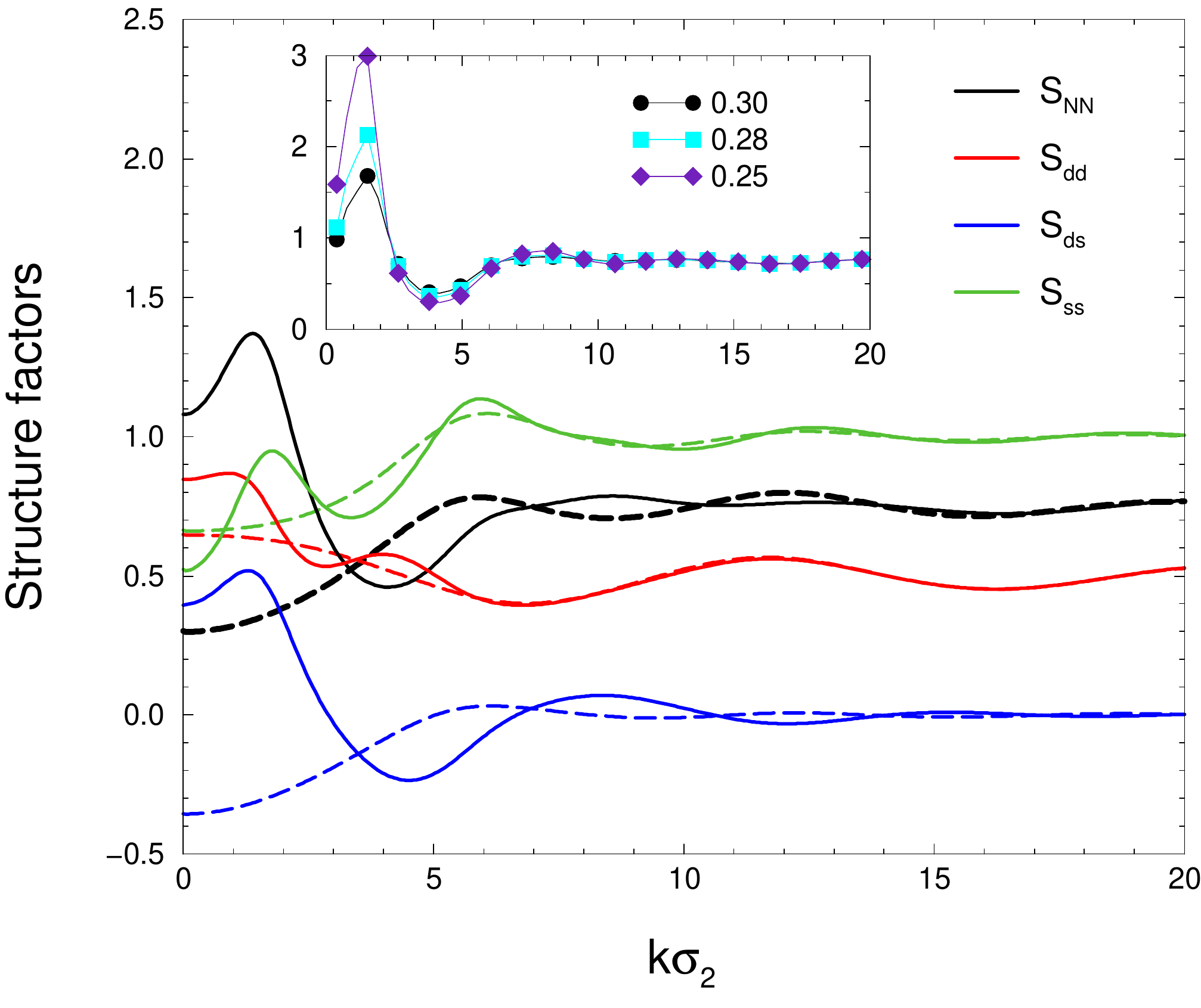} 
\end{tabular}
\caption{RISM $S_{\rm NN}(k)$ resolved into various contributions,
for $T^\ast=0.30$ and $\chi_{\rm s}=0.5$ (full lines); 
for comparison, we also report the predictions for $T=\infty$ (dashed lines),
i.e. the same curves as shown in Fig.~\ref{fig:highT}B. 
Inset: MC $S_{\rm NN}(k)$ for low temperatures (in the legend).}
\label{fig:snn}
\end{center}
\end{figure}
%%%%%%%%%%%%%%%%%%%%%%%%%%%%%%%%%%%%%%%%%%%%%%%%%%%%%%%%
%%%%%%%%%%%%%%%%%%%%%%%%%%%%%%%%%%%%%%%%%%%%%%%%%%%%%%%%
\begin{figure}[!t]
\begin{center}
\begin{tabular}{c}
\includegraphics[width=6.8cm,angle=-90]{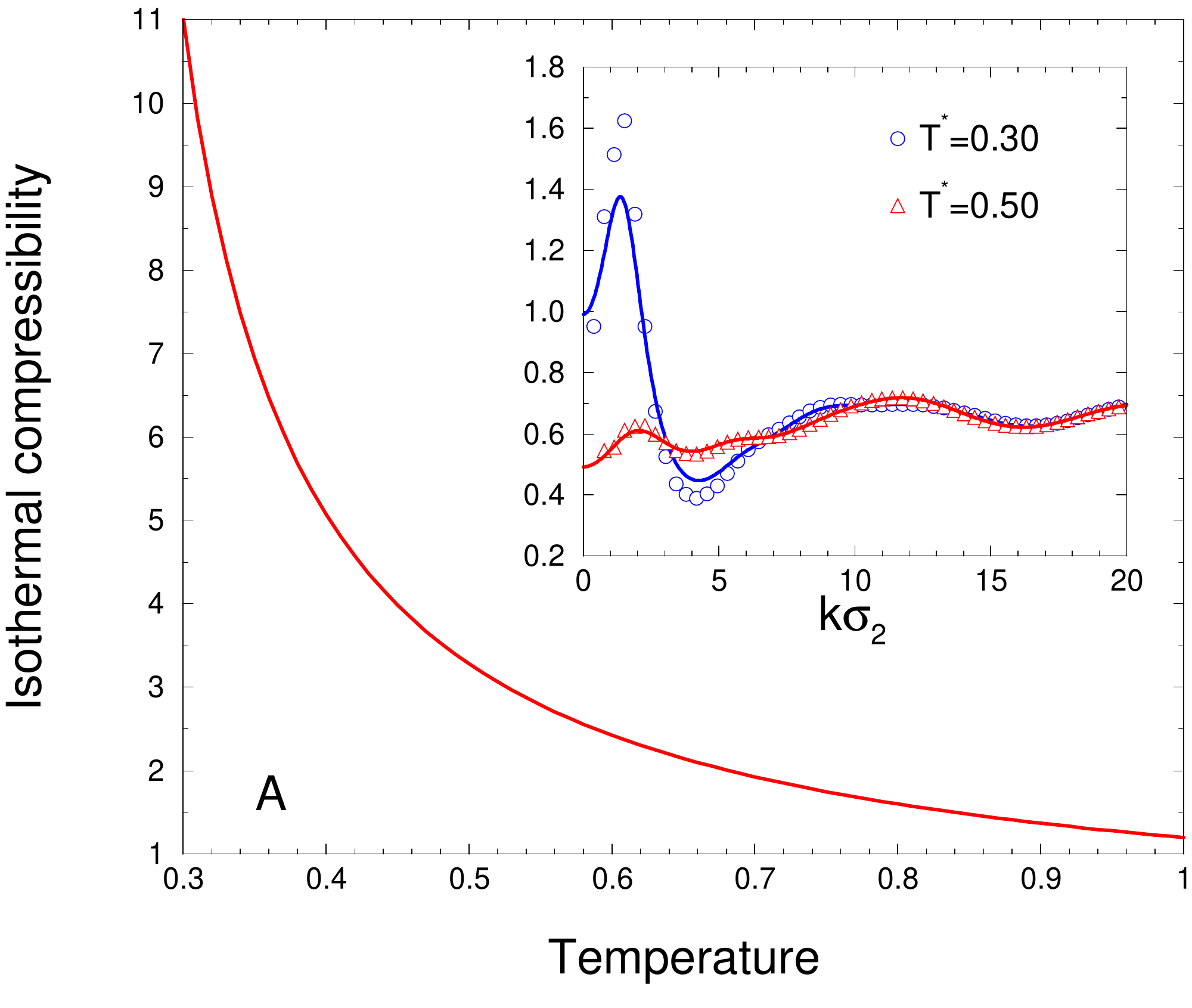} \\ 
\includegraphics[width=6.8cm,angle=-90]{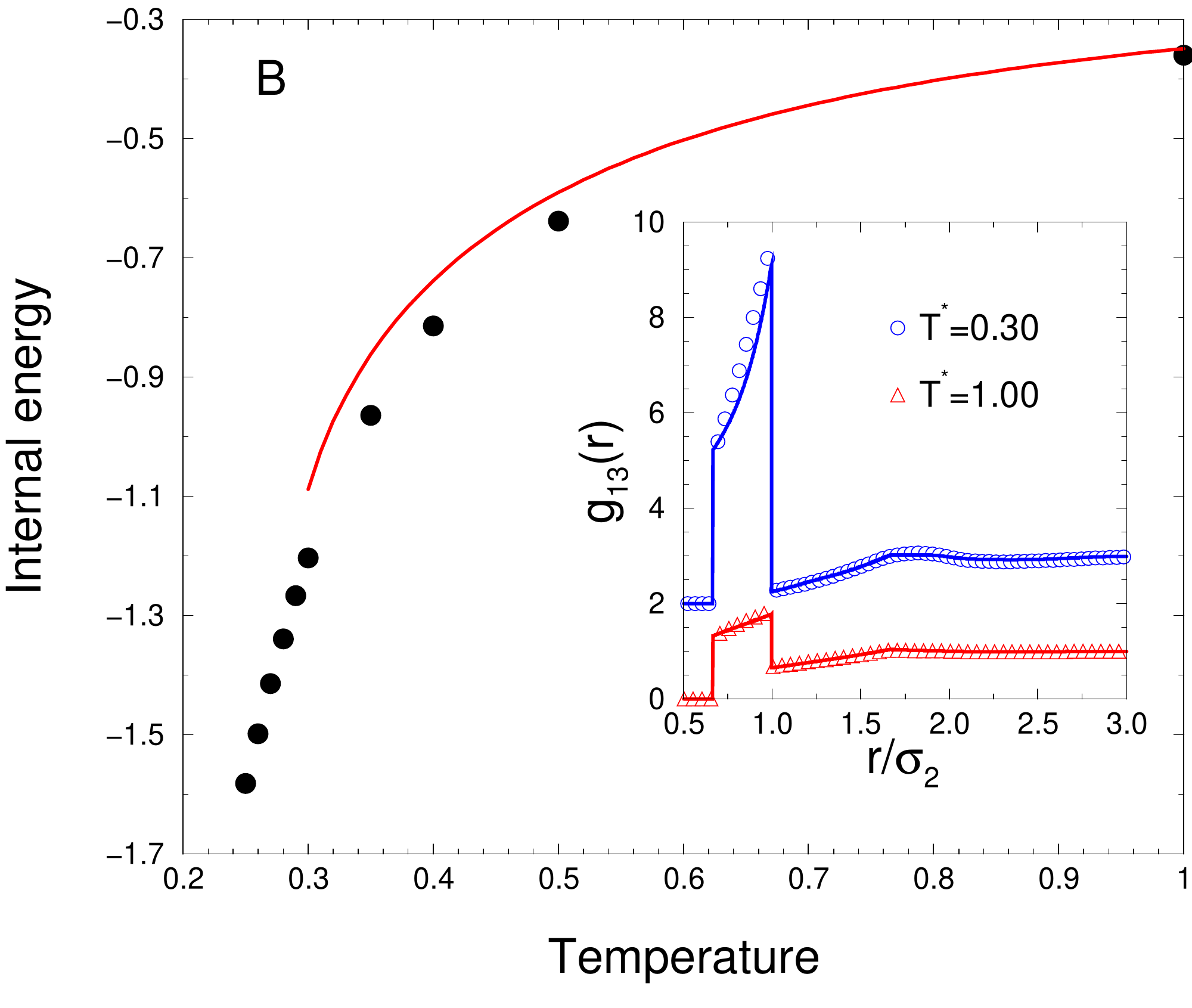} 
\end{tabular}
\caption{A: RISM isothermal compressibility at equimolarity
as a function of temperature.
Inset: the r.h.s. of (\ref{eq:comp}) according to RISM (lines) and MC (symbols),
extended over the whole $k$ range for two different temperatures (see the legend).
B: RISM (lines) and MC (symbols) energies, for the same thermodynamic conditions as before.
Inset: $g_{13}(r)$ according to RISM and MC for two different temperatures
(see the legend); for the sake of clarity, $g_{13}(r)$ for $T^\ast=0.30$ has been
vertically shifted by two.}
\label{fig:thermo}
\end{center}
\end{figure}
%%%%%%%%%%%%%%%%%%%%%%%%%%%%%%%%%%%%%%%%%%%%%%%%%%%%%%%%

Concerning thermodynamic properties, the compressibility of the fluid
as computed by RISM is plotted in Fig.~\ref{fig:thermo}A as a function
of the temperature. Looking for an assessment of this quantity by MC,
we might think that an accurate check of RISM data cannot be carried out,
due to the well known problem to extrapolate the $k\rightarrow 0$
behavior of structure factors from finite-size simulations.
In an attempt to remedy this difficulty, rather than extrapolating the
individual MC structure factors down to $k=0$, we have computed
the r.h.s. of Eq.~(\ref{eq:comp}) over the whole $k$ range, see the inset of
Fig.~\ref{fig:thermo}A. Here we note
that RISM and MC substantially agree
near $k=0$, implying that the compressibility estimate provided by RISM
can be deemed as good.
From the same inset we see that the only $k$ region where RISM
deviates from MC is again near the low-$k$ peak.
Turning back to the main panel, we observe 
that the compressibility
moderately increases as the temperature is lowered at fixed concentration and density,
with no hints at a diverging trend near the ultimate threshold of
convergence of the RISM algorithm; hence, we can exclude
any propensity of the mixture towards a macroscopic phase separation.
This conclusion is clearly consistent with the presence of a low-$k$ peak
in $S_{\rm NN}(k)$, possibly evolving --- at lower temperatures ---
to a divergence at finite $k$, like in other fluids with local inhomogeneities.

The internal energy per particle is shown in Fig.~\ref{fig:thermo}B.
We see from this picture that, within its operating range, the RISM theory
closely follows MC results. This good
agreement reflects the accurate
theoretical prediction of $g_{13}(r)$, as exemplified in the inset of panel B
for two different temperatures.
In the main panel, we see a progressive decrease of 
the internal energy on cooling,
with the parabolic trend for $T^\ast \gtrsim 0.30$ replaced by a 
roughly linear behavior below this threshold.

%%%%%%%%%%%%%%%%%%%%%%%%%%%%%%%%%%%%%%%%%%%%%%%%%%%%%%%%
\begin{figure}
\begin{center}
\begin{tabular}{c}
\includegraphics[width=6.5cm,angle=-90]{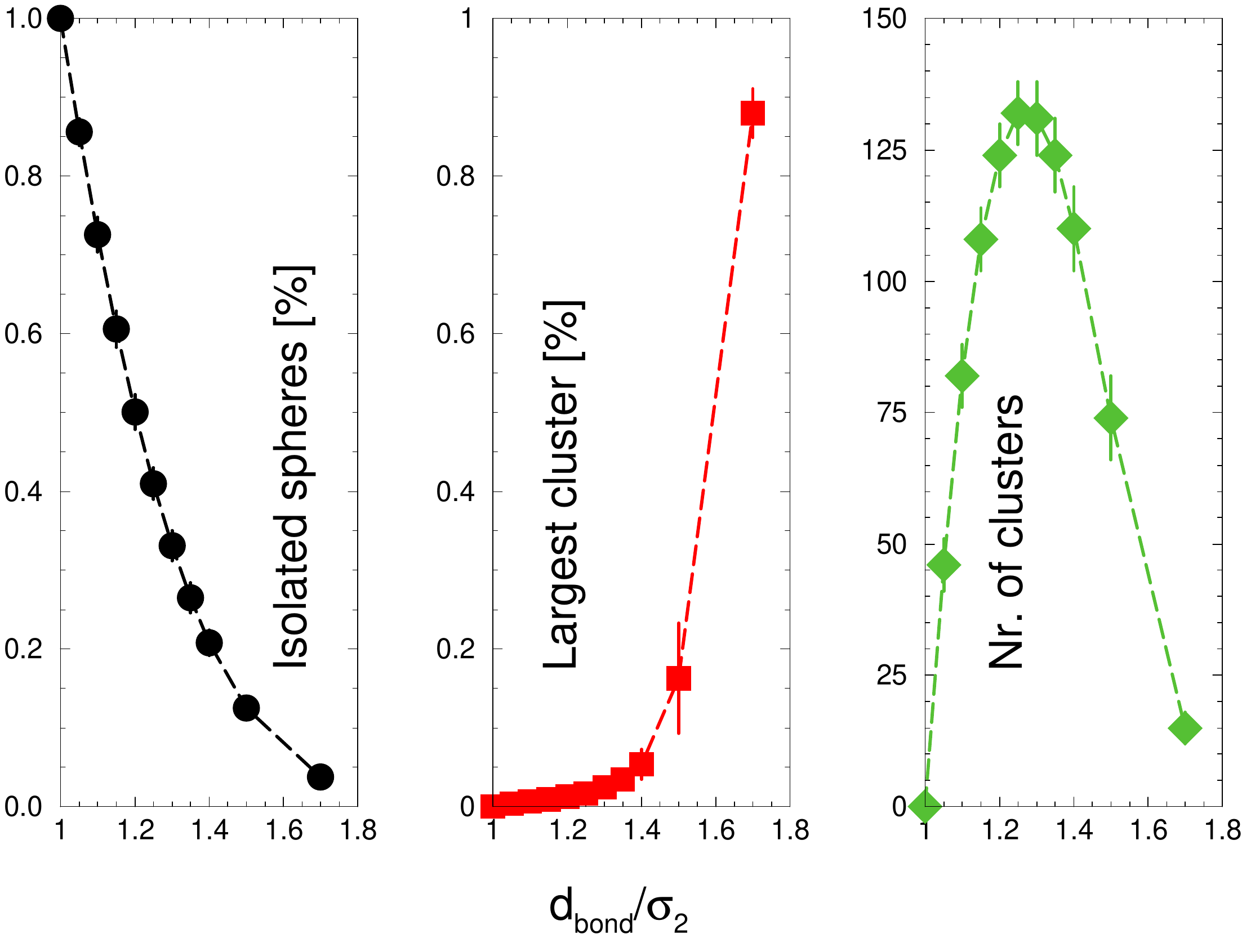}
\end{tabular}
\caption{MC data for the fraction of isolated spheres (left), the
largest cluster size (center), and the total number of clusters (right),
as a function of $d_{\rm bond}$ for $\chi_{\rm s}=0.5$ and infinite temperature.
Average values per single configuration, as computed over about 1000 configurations
uniformly distributed along the last part of the MC trajectory, are reported
with the corresponding statistical uncertainties. Lines are guides to the eye.
}\label{fig:averclus-T100}
\end{center}
\end{figure}
%%%%%%%%%%%%%%%%%%%%%%%%%%%%%%%%%%%%%%%%%%%%%%%%%%%%%%%%

Simulation offers the possibility to relate the emerging structural and thermodynamic features
with the local arrangement of particles in the fluid. This opportunity prompts us to carry out
an extended analysis of aggregation properties, based on the microscopic configurations
generated by MC.
In order to characterize the aggregation properties of the mixture, 
a crucial issue is the definition of the ``bond distance'' $d_{\rm bond}$,
i.e. the distance within which two particles can be considered as bonded together.
As a preliminary consideration, since spheres are non-interacting beyond the hard core,
it is not straightforward to associate $d_{\rm bond}$ with the range of
attractive interactions, as is commonly assumed in studies of SALR fluids. 
In our previous analysis,~\cite{Prestipino:19} two spheres were considered as
bonded together when their distance is smaller
than $\sigma_3+3\sigma_1 \equiv 2\sigma_2$,
which represents --- according to Eq.~(\ref{eq:pot}) --- the maximum distance at which two
spheres can experience a mutual attraction mediated by a small monomer placed in the middle. 
However, this choice would lead to a too large value of
the bonding distance, given the relatively high density of our sample.
Indeed, we have checked that $d_{\rm bond}=2$ 
returns the unphysical outcome that all spheres belong to a single aggregate
spanning the whole simulation box, even in the high temperature regime where
attractive interactions are altogether absent.
The same conclusion follows if we adopt another common choice,
which is to set $d_{\rm bond}$ equal to the position of
the first minimum in $g_{33}(r)$ ($\approx 1.8$).
At the opposite end, if we were to use a too much restrictive criterion,
for instance $d_{\rm bond}=1.1$, then we would conclude 
that aggregation is practically absent even at very low temperature,
at variance with the picture emerging from the structural analysis.

Seeking for an optimal bond distance in the interval between $1.1$ and $1.8$,
we turn back to our high-temperature sample. We have computed a few
aggregation properties as a function of $d_{\rm bond}$, based on the available
microscopic MC configurations. As reasonably expected, see
Fig.~\ref{fig:averclus-T100} (left panel), the average fraction
of isolated hard spheres per single configuration 
constantly decreases as $d_{\rm bond}$ grows, whereas the opposite
occurs for the size of the largest cluster (middle panel).
On the other hand, in the right panel we see that the total number of aggregates
in the sample, whatever their size, shows a non-monotonic trend 
as a function of $d_{\rm bond}$, with a maximum around $1.25$.
This outcome can be explained by observing that the number of aggregates
initially grows with $d_{\rm bond}$, since more and more spheres
are allowed to join to each other at the expenses of isolated particles. 
However, beyond a certain threshold aggregates coalesce
to form larger clusters and this progressively lowers their number.
At this point, it is reasonable to take $d_{\rm bond}\approx 1.25$,
namely to assume as bond distance the value maximizing,
at high temperature, the pool of possible outcomes in terms
of number of aggregates. By this choice, we ensure
that the counting of aggregates is the least sensitive to small variations
of $d_{\rm bond}$ from the point of maximum.
%%%%%%%%%%%%%%%%%%%%%%%%%%%%%%%%%%%%%%%%%%%%%%%%%%%%%%%%
\begin{figure}[!t]
\begin{center}
\begin{tabular}{c}
\includegraphics[width=6.5cm,angle=-90]{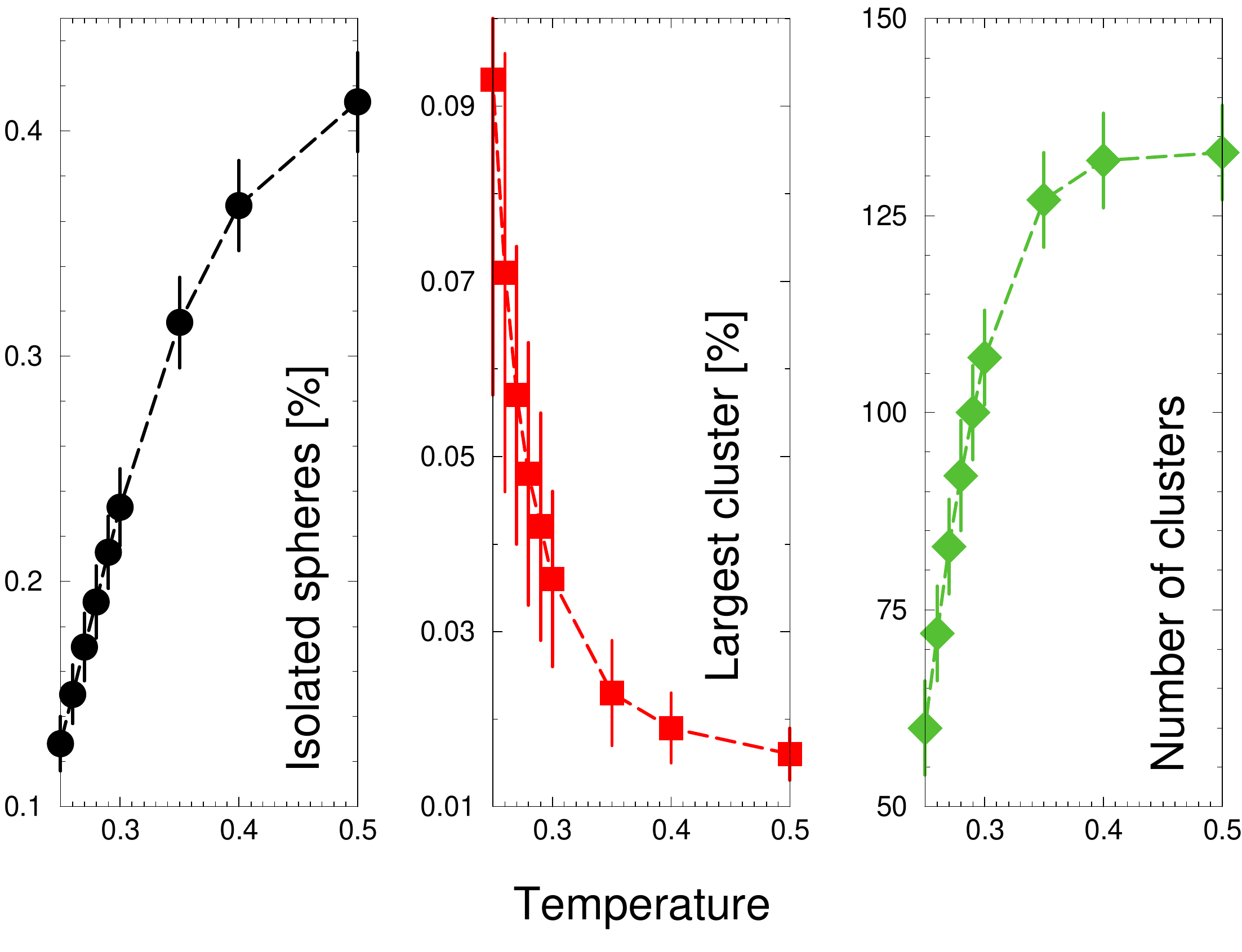} 
\end{tabular}
\caption{
Same properties as in the previous figure, but for fixed $d_{\rm bond}=1.25\sigma_2$,
plotted as functions of temperature for $\chi_{\rm s}=0.5$.
}\label{fig:averclus-d125}
\end{center}
\end{figure}
%%%%%%%%%%%%%%%%%%%%%%%%%%%%%%%%%%%%%%%%%%%%%%%%%%%%%%%%
Slightly anticipating the discussion of
Fig.~\ref{fig:csd}, the cluster-size distribution associated
with $d_{\rm bond}=1.25$ is characterized at high temperature
by a sharp decay with size, with practically zero
probability to find more than 10-12 particles connected together.
In other words, even in the absence of attraction, small aggregates statistically 
form and dissolve following the natural MC evolution of the mixture. 
It is clear that, according to the proposed criterion, $d_{\rm bond}$ 
needs to be calculated every time the density of hard spheres is changed,
for example when we change the relative concentration at fixed total density.
%%%%%%%%%%%%%%%%%%%%%%%%%%%%%%%%%%%%%%%%%%%%%%%%%%%%%%%%
\begin{figure}
\begin{center}
\begin{tabular}{c}
\includegraphics[width=6.5cm,angle=-90]{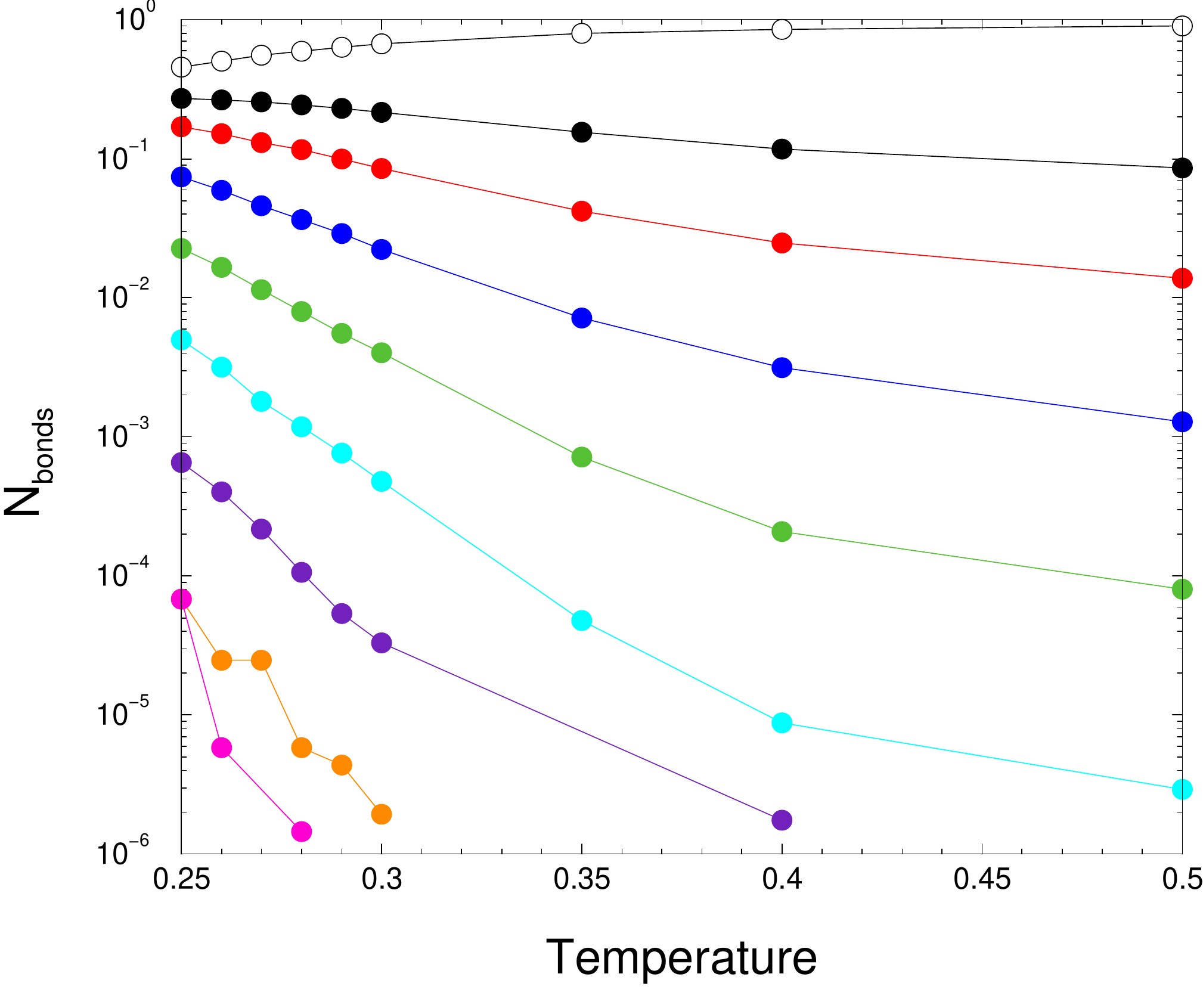} \\
\end{tabular}
\caption{{Number of bonds per sphere plotted as a function of temperature
for $\chi_{\rm s}=0.5$. From top to bottom, $N_{\rm bonds}=0, 1,$ \ldots, 8.   
}}\label{fig:bonds}
\end{center}
\end{figure}
%%%%%%%%%%%%%%%%%%%%%%%%%%%%%%%%%%%%%%%%%%%%%%%%%%%%%%%%

Choosing $d_{\rm bond}=1.25$, we have studied the same three properties
in Fig.~\ref{fig:averclus-T100} as a function of temperature, 
in order to detect, from a microscopic viewpoint, the tendency of the system 
to form aggregates. This analysis is reported in Fig.~\ref{fig:averclus-d125}.
We see that the number of isolated spheres (left panel) drops upon cooling,
until, for $T^\ast =0.25$, only $\approx 10$\% of the spheres are non-bonded.
At the same time, small aggregates progressively coalesce into larger units;
as a result, the total number of clusters (right panel) rapidly decreases,
while the size of the largest cluster (middle panel) increases, until it contains
about 10\% of all spheres in the mixture.

As the temperature is lowered, spheres reorganize themselves in space
in such a way that their local environment becomes progressively denser.
This can be seen from Fig.~\ref{fig:bonds}, where the number of bonds
per sphere, $N_{\rm bonds}$, is reported as a function of temperature.
We see that the fraction of spheres engaged in one or more bonds
steadily increases at the expenses of isolated spheres.
In particular, for $T^*=0.25$ more than 50\% of the spheres are bonded
to at least another sphere. Interestingly enough, for $T^\ast=0.30$
we see the first occurrence of a sphere with seven neighbors;
spheres involved in eight bonds first show up for $T^\ast=0.28$.

%%%%%%%%%%%%%%%%%%%%%%%%%%%%%%%%%%%%%%%%%%%%%%%%%%%%%%%%
\begin{figure}[!b]
\begin{center}
\begin{tabular}{c}
\includegraphics[width=6.2cm,angle=-90]{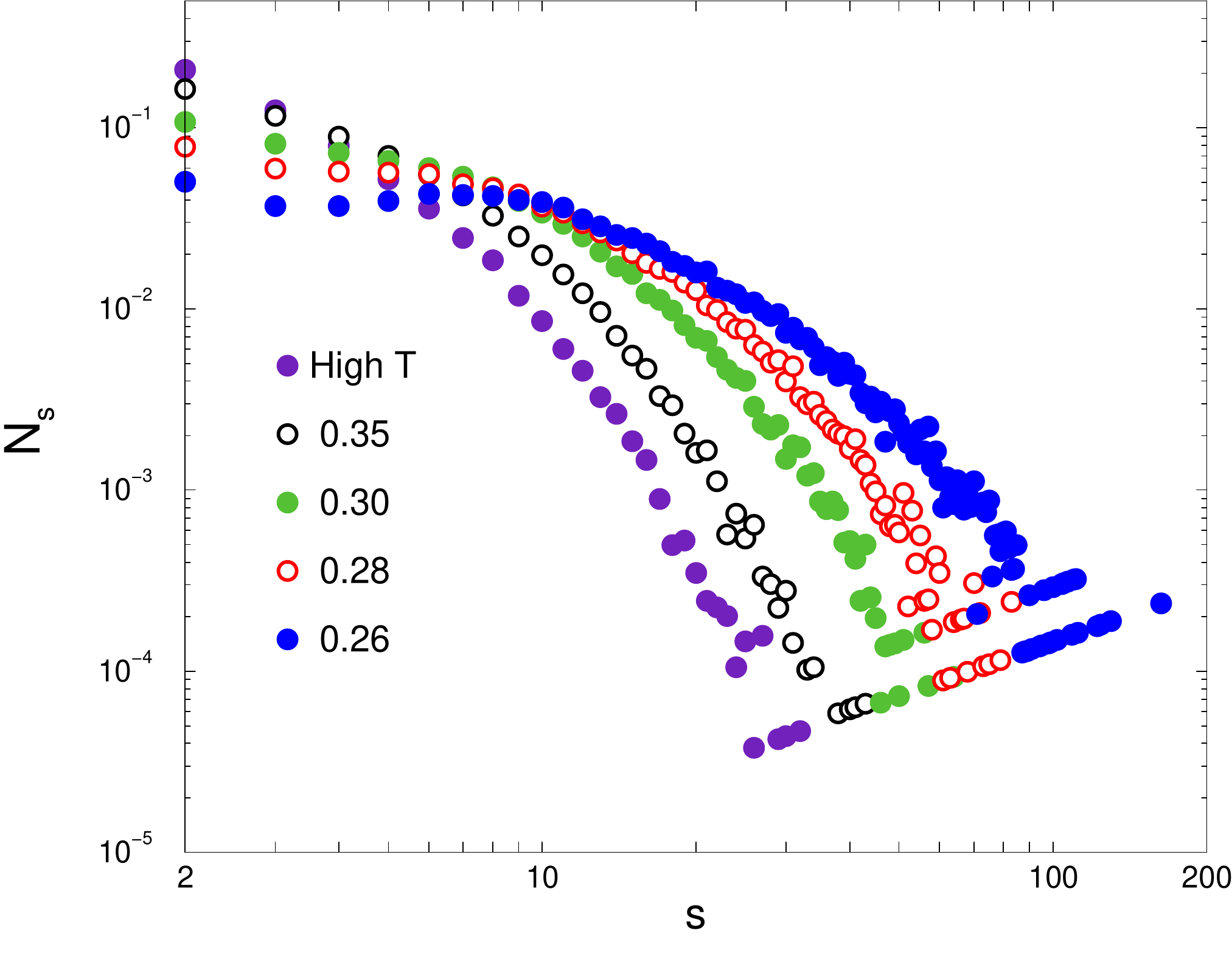} \\
\end{tabular}
\caption{{Cluster-size distribution of hard spheres for a few selected temperatures
(see the legend) and for $\chi_{\rm s}=0.5$.
}}\label{fig:csd}
\end{center}
\end{figure}
%%%%%%%%%%%%%%%%%%%%%%%%%%%%%%%%%%%%%%%%%%%%%%%%%%%%%%%%

A more detailed picture of the microscopic arrangement of hard spheres
is provided by the cluster-size distribution [Eq.~(\ref{eq:csd})],
reported in Fig.~\ref{fig:csd} for $\chi_{\rm s}=0.5$. We note that $N(s)$ 
undergoes a sharp decay for $T^\ast=0.35$.
In the range $T^\ast=0.30-0.27$ the decay is less sharp
and an inflection point occurs, until eventually,
for $T^\ast=0.26$, we find for the first time the occurrence
of a local peak in the distribution, signaling the presence of clusters
(or, better, ``aggregates'') preferentially composed by $\approx 10$ spheres
(over a pool of 686 hard spheres). At this low temperature most aggregates
are formed on average by ten particles or so, whereas the largest aggregate
involves about 70 spheres (see also the middle panel of Fig.~\ref{fig:bonds});
only occasionally aggregates with more than 100 spheres are seen.

%%%%%%%%%%%%%%%%%%%%%%%%%%%%%%%%%%%%%%%%%%%%%%%%%%%%%%%%
\begin{figure}[!b]
\begin{center}
\begin{tabular}{c}
\vspace*{-0.5cm}
\includegraphics[width=9.0cm,angle=0]{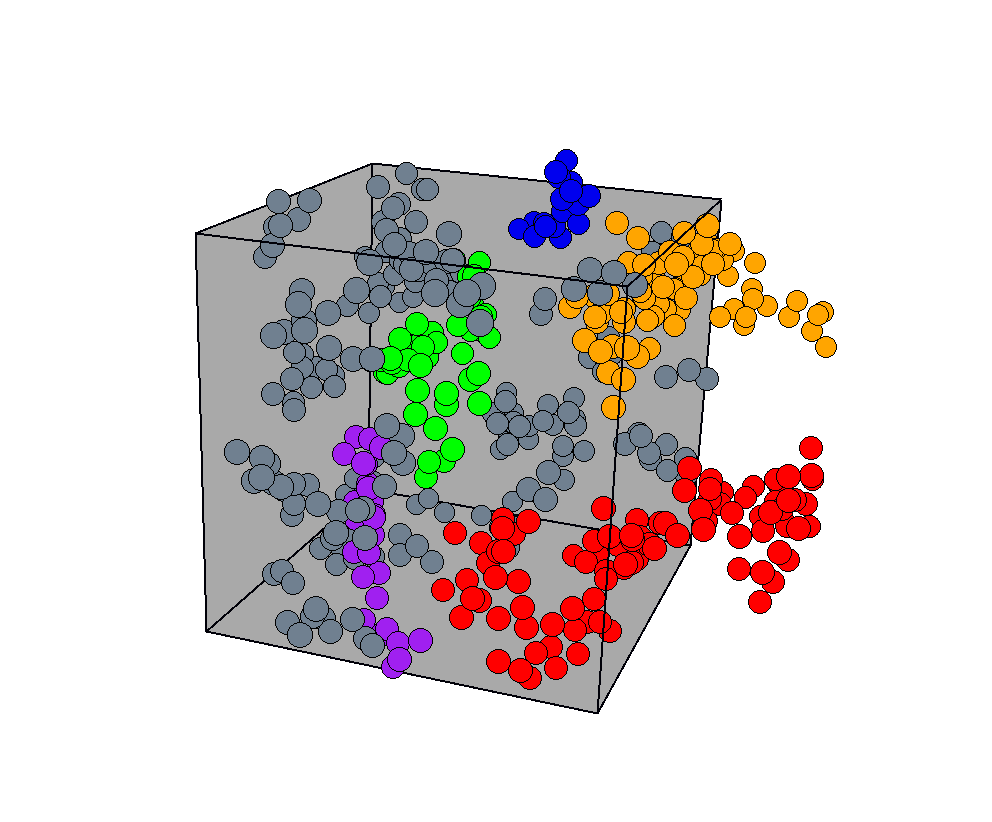} \\
\includegraphics[width=7.0cm,angle=0]{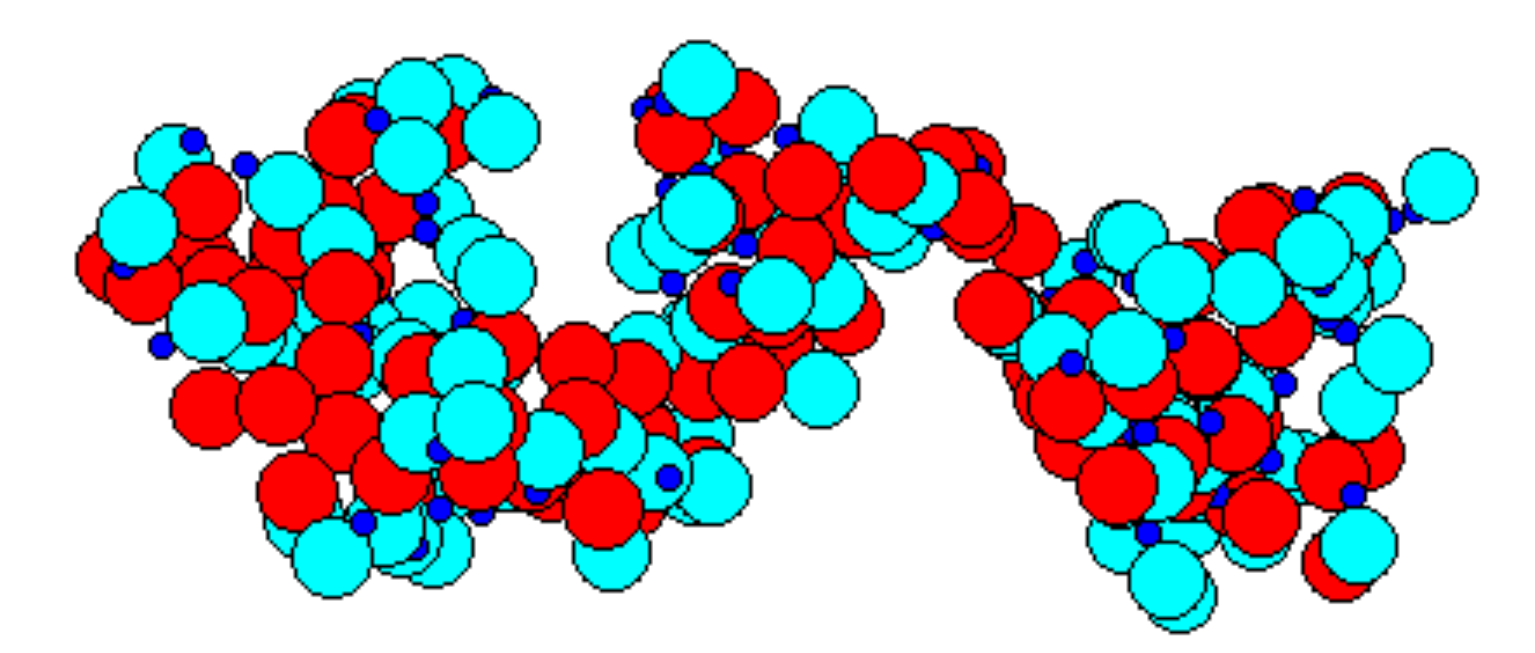}
\end{tabular}
\caption{{Top: microscopic picture of an equilibrium configuration
for $T^\ast=0.25$ and $\chi_{\rm s}=0.5$.
For a better visualization, neither dimers nor isolated spheres
($\approx 80$) are shown.
Most aggregates are formed by less than 10 spheres (in grey).
The largest cluster (in red) is formed
by $\approx 90$ spheres, followed in descending order by
aggregates with 75 (orange), 42 (green), 26 (purple), and 17 (blue)
spheres. Bottom: detail of the largest aggregate, with spheres (red)
drawn together with their neighboring dimers; according to the
color convention of Fig.~\ref{fig:models}, large monomers
are colored in cyan, whereas small monomers in blue.
}}\label{fig:snap}
\end{center}
\end{figure}
%%%%%%%%%%%%%%%%%%%%%%%%%%%%%%%%%%%%%%%%%%%%%%%%%%%%%%%%
At this point, we can resume all structural (Figs.~\ref{fig:lowT} and~\ref{fig:snn})
and microscopic information (Fig.~\ref{fig:csd}) gathered so far, to attempt a unified picture
of the phase behavior of the fluid mixture upon cooling. Again, our reasoning
is inspired by similar studies of one-component SALR fluids~\cite{Godfrin:13,Godfrin:14}.
At very high temperature, no low-$k$ peak is found in the structure factors;
in this case the fluid is fully homogeneous and the cluster-size distribution
is characterized by a sharp decay of the probability associated with the formation of
clusters of increasing size --- always allowed by random fluctuations,
even in the absence of any attraction.
As the temperature is lowered, a low-$k$ peak first appears and then steadily grows
in height $(T^\ast=1.00-0.27)$; correspondingly, the fluid displays 
local inhomogeneities
that are progressively more marked; $N(s)$ is still characterized by a monotonic decay
and the fluid is said to exhibit ``intermediate-range order'';~\cite{Godfrin:13,Godfrin:14}
as soon as a maximum appears in $N(s)$ (here for $T^\ast=0.26$), a more definite ``clustered state''
is entered, with the location of the maximum signaling the typical size of clusters.

The analysis presented so far already demonstrates the appearance
of aggregates in the system at low temperature.
However, it gives no information about their structure.
Even though a geometric characterization of the aggregates is out of the scope
of the present paper, some information can be got from a visual inspection
of microscopic configurations. To exemplify, we show in Fig.~\ref{fig:snap}
a typical snapshot of the equilibrated sample for $T^\ast=0.25$.
To avoid the breaking of aggregates at the boundary of the box,
we have first computed the center of mass of each aggregate, according
to the procedure devised by Bai and Breen;~\cite{Bai:08}
then, for each particle in a given aggregate we have considered its
closest replica to the center of mass. We see that a large aggregate
generally takes an elongated, relatively open conformation, with no
appreciable branching. This evidence can be explained by looking
at the detailed structure of the largest aggregate, shown in the bottom
panel of Fig.~\ref{fig:snap}, where each sphere is plotted together with
its neighboring dimers, which are those for which the 1-3 distance
falls within the attraction range of Eq.~(\ref{eq:pot}).
As is clear, spheres are joined together by small monomers
placed in the interstitial spaces between them; large monomers
form a layer around spheres, with a double effect; on one hand,
this inert coating prevents the possibility for spheres to come closer to each other,
thus giving aggregates a more compact, globular form that is more typically
associated with the common concept of ``cluster'' of particles.
Secondly, the same hindrance effect prevents aggregates from coalescing
together, i.e. forming a single or a few large droplets
which would typically anticipate a full phase separation.
We do not follow the ultimate fate of our fluid at still lower temperatures,
where it eventually becomes structurally arrested.~\cite{Prestipino:19}

%%%%%%%%%%%%%%%%%%%%%%%%%%%%%%%%%%%%%%%%%%%%%%%%%%%%%%%
\begin{figure}
\begin{center}
\begin{tabular}{c}
\includegraphics[width=4.1cm,angle=-90]{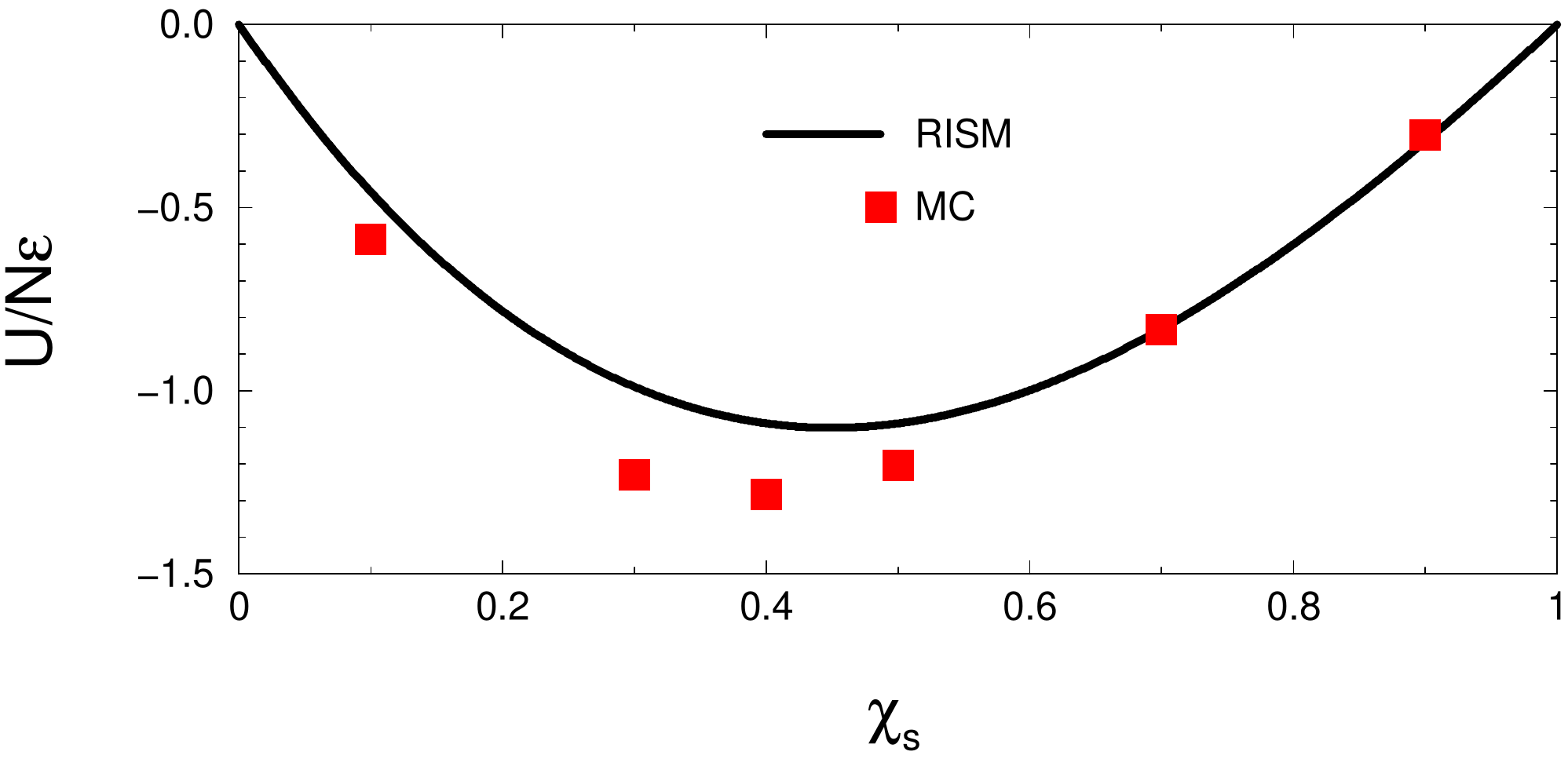}
\end{tabular}
\caption{
RISM (lines) and MC (symbols) internal energies per particle, plotted as functions
of the concentration for $T^\ast=0.30$ and $\rho^\ast=0.30$.
}\label{fig:energy-chi}
\end{center}
\end{figure}
%%%%%%%%%%%%%%%%%%%%%%%%%%%%%%%%%%%%%%%%%%%%%%%%%%%%%%%%

Having broadly characterized the properties of the mixture under
equimolar conditions, the last part of our study is devoted
to the analysis of other values of the sphere concentration $\chi_{\rm s}$,
while still holding the overall density fixed at $\rho^\ast=0.30$.
In this regard, we observe that changes in the concentration of spheres
do not appreciably affect the overall convergence properties of the RISM algorithm;
in particular, the lowest temperature that can be attained by RISM is still $T^\ast=0.30$.
RISM predictions will be discussed near this low-temperature threshold only.

Starting from thermodynamics, MC and RISM energies per particle
are shown in Fig.~\ref{fig:energy-chi} for various concentration values.
Both schemes substantially agree in signaling that the strongest cohesion
between spheres and dimers is attained when the former are
slightly less numerous than the latter (i.e. for $\chi_{\rm s} \approx 0.3-0.4$).
More precisely, RISM predicts a minimum energy for $\chi_{\rm s}=0.45$,
only slightly shifted from the MC datum $\chi_{\rm s}=0.40$.
Overall, RISM agrees with MC all over the concentration range,
though the RISM minimum is more rounded and slightly overestimated.

%%%%%%%%%%%%%%%%%%%%%%%%%%%%%%%%%%%%%%%%%%%%%%%%%%%%%%%%
\begin{figure}[!t]
\begin{center}
\begin{tabular}{c}
\includegraphics[width=6.8cm,angle=-90]{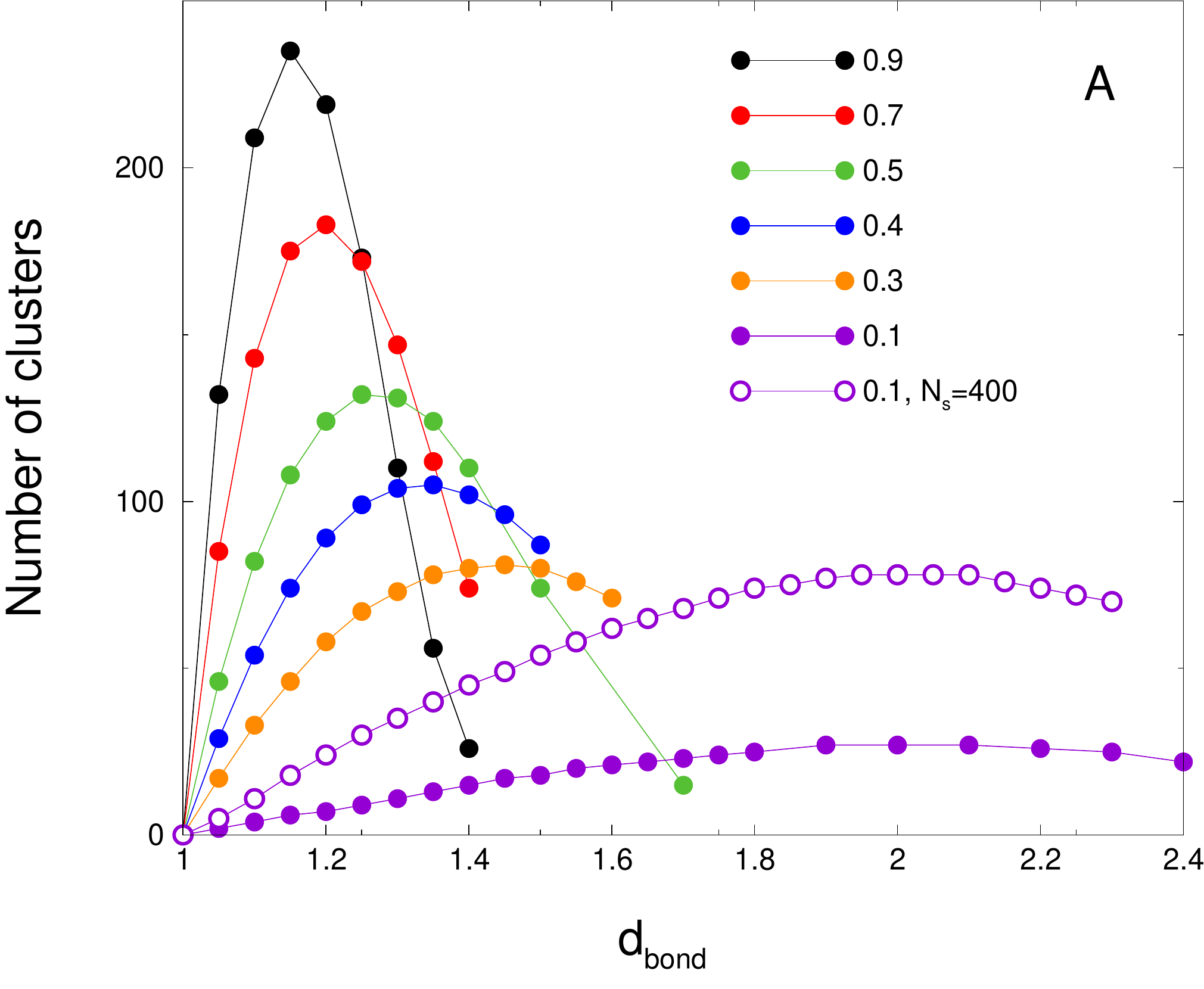} \\
\includegraphics[width=6.8cm,angle=-90]{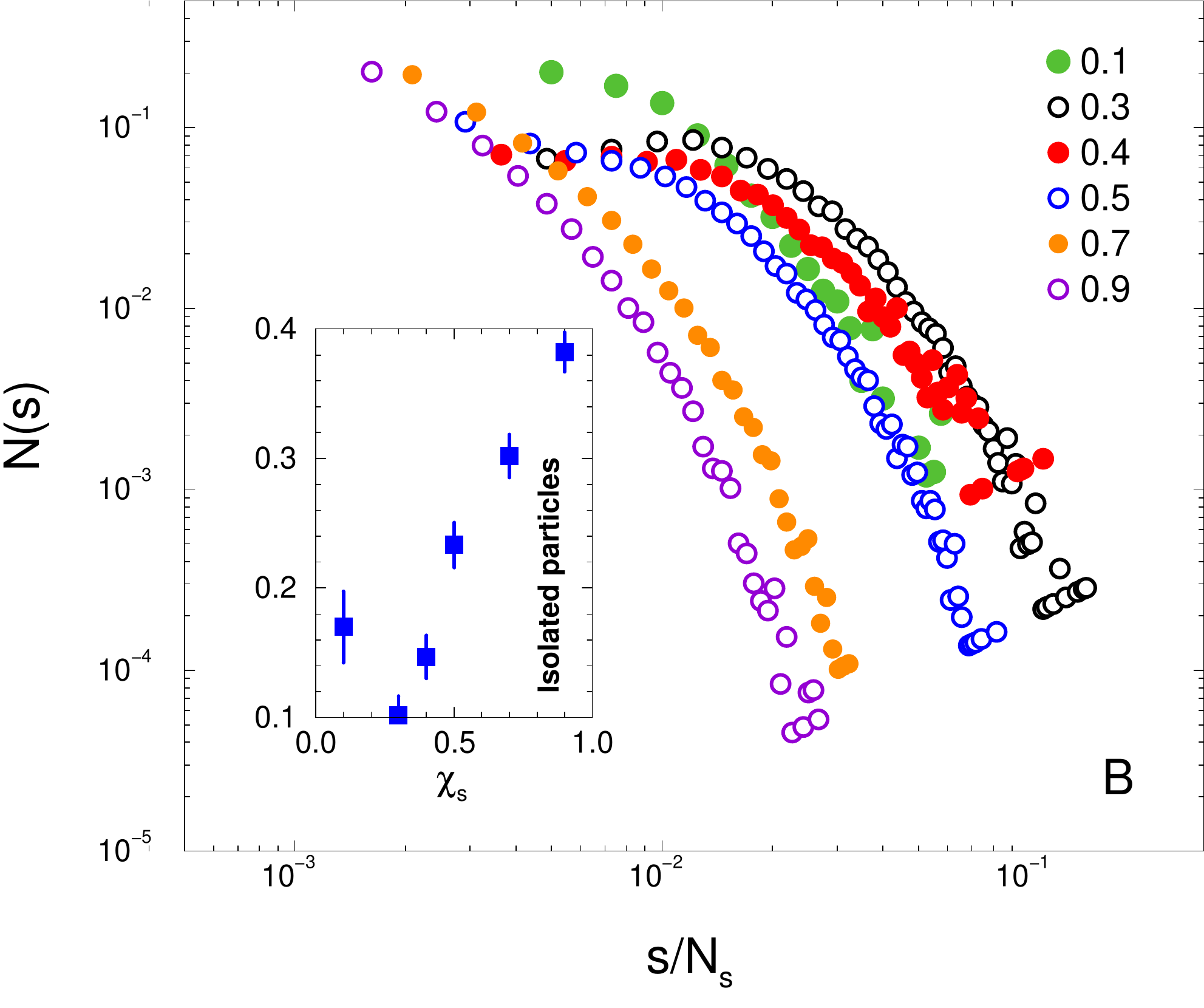}
\end{tabular}
\caption{A: Total number of clusters plotted as a function of
$d_{\rm bond}$ for decreasing concentration of spheres (see the legend)
at infinite temperature. Open circles refer to a sample composed by a total of
$N=4000$ particles. B: cluster-size distribution for a few $\chi_{\rm s}$ values
(see the legend), for $T^\ast=0.30$. For a better comparison, the
cluster size $s$ has been normalized by the number of spheres
at the given concentration. Inset: fraction of isolated particles vs $\chi_{\rm s}$.
\label{fig:averclus-chi}}
\end{center}
\end{figure}
%%%%%%%%%%%%%%%%%%%%%%%%%%%%%%%%%%%%%%%%%%%%%%%%%%%%%%%%

In order to link the energy information with the microscopic arrangement
of particles, we have preliminarily determined the value of $d_{\rm bond}$
for a number of concentration values, by repeating the same calculation
previously described for $\chi_{\rm s}=0.5$ (where $d_{\rm bond}=1.25$).
Results are shown in Fig.~\ref{fig:averclus-chi}A.
When the number of spheres is large, e.g., for $\chi_{\rm s}=0.9$,
the bond distance is reduced to $d_{\rm bond}=1.15$, which is coherent
with the general premises of our method: 
at higher densities, spheres are closer together; hence 
a maximum in the number of aggregates falls at shorter $d_{\rm bond}$.
As the concentration of spheres decreases, $d_{\rm bond}$ increases until
it attains the upper limit of $d_{\rm bond}=2$ for $\chi_{\rm s}=0.1$.
In this limit, the contact distance becomes equal to the interaction distance.
Simulations with $4000$ particles for $\chi_{\rm s}=0.1$ (open circles in
Fig.~\ref{fig:averclus-chi}A) show that, despite a systematic increase
in the number of clusters, the general shape of the curve and the location
of the maximum are left unchanged.

\begin{figure}[t]
\begin{center}
\begin{tabular}{c}
\includegraphics[width=6.8cm,angle=-90]{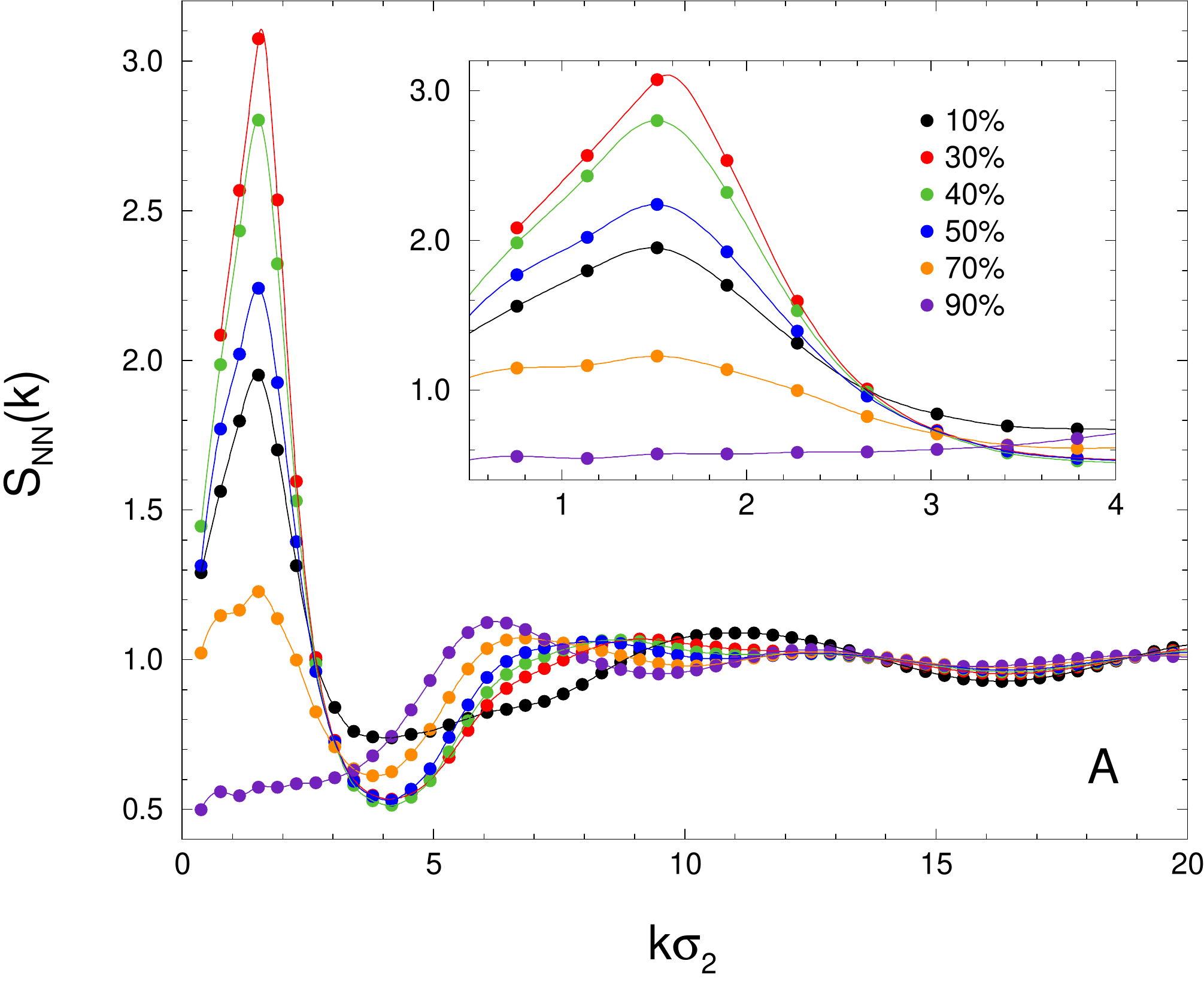}  \\
\includegraphics[width=6.8cm,angle=-90]{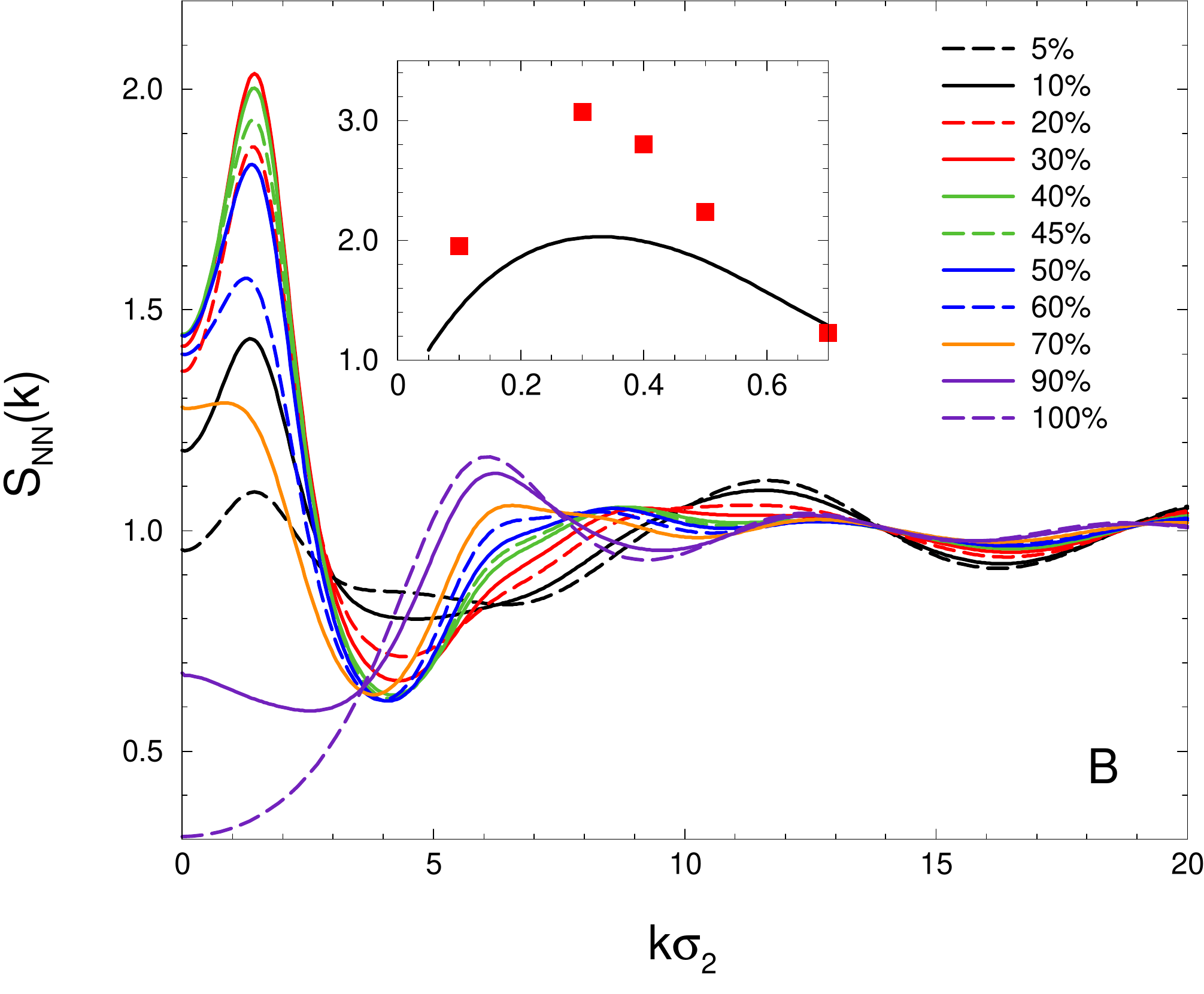}
\end{tabular}
\caption{A: MC $S_{\rm NN}(k)$ for various $\chi_{\rm s}$ values
(see the legends) and $T^\ast=0.30$. Inset: magnification of the low-$k$ peak.
B: same property, as predicted by RISM. Inset: height of the low-$k$ peak
according to MC (squares) and RISM (line).
\label{fig:structure-chi}}
\end{center}
\end{figure}

Once the value of $d_{\rm bond}$ has been set for every $\chi_{\rm s}$,
we can compute the cluster-size distribution for $T^\ast=0.30$
(see Fig.~\ref{fig:averclus-chi}B). For a better comparison between
the various concentrations, the cluster size has been normalized
to the number of spheres present in the mixture at the given $\chi_{s}$.
The overall appearance of $N(s)$ complies with the picture emerging
from the behavior of internal energy: when spheres dominate
(e.g., for $\chi_{\rm s}=0.9$), the few dimers present are insufficient to provide
the ``glue'' necessary to join all the spheres together, thus leading to a sharp decay
of $N(s)$, which is tantamount to a large fraction of isolated spheres
($\approx 40$\%, see the inset). Things change when dimers
are progressively added. When $\chi_{\rm s}$ drops to $0.5$, still
25-30\% of the spheres are isolated and $N(s)$ is shifted towards larger $s$.
At the opposite end ($\chi_{\rm s}=0.1$), dimers are so many
that the number of isolated spheres is reduced to $\approx 17$\%; in this case,
$N(s)$ extends to larger $s$ in comparison with $\chi_{\rm s}=0.7$ or 0.9.
In between, the largest number of contacts between spheres and dimers is
obtained for $\chi_{\rm s}=0.3-0.4$, where only a few spheres
(roughly 10\% of the total) are not involved in the aggregation process
and $N(s)$ extends to the largest $s$ values of all.
All indications now agree to conclude that the optimal conditions
for the stabilization of the mixture are met for $\chi_{\rm s}=0.3-0.4$.
Looking at Fig.~\ref{fig:averclus-chi}B, we see that $N(s)$ shows a monotonic
decay for $\chi_{\rm s}=0.1$ and in the interval $0.5-0.9$;
as discussed before, under these conditions the fluid exhibits intermediate-range order.
On the other hand, $N(s)$ shows a plateau for $\chi_{\rm s}=0.4$, which for
$\chi_{\rm s}=0.3$ gives way to a shallow local maximum around $s\approx 5$,
indicating the preferential development of aggregates with that size.
In both cases, the mixture is closer in temperature to conditions favorable to
the development of a clustered state, similarly as found for $\chi_{\rm s}=0.5$
and $T^\ast=0.26$.

Finally, the total structure factor $S_{\rm NN}(k)$ is reported in
Fig.~\ref{fig:structure-chi} for $T^\ast=0.30$, where results from 
MC simulation (A) and RISM theory (B) are shown for various
concentrations of spheres.
As for MC, a low-$k$ peak is practically absent for $\chi_{\rm s}=0.9$
and only hardly seen for $\chi_{\rm s}=0.7$ (see also the inset);
a more distinct peak is visible for $\chi_{\rm s}=0.1$ and 0.5,
whereas the low-$k$ peak reaches its maximum height
for intermediate concentrations, i.e. for $\chi_{\rm s}=0.3,$ and 0.4. Hence,
a gratifying agreement is found between the shape of the low-$k$ peak
and the information obtained from the previous cluster analysis.
All this witnesses the good quality of the procedure followed
in the definition of $d_{\rm bond}$, since a coherent indication
is returned from two independent (microscopic and structural) sources.
We may appreciate in Fig.~\ref{fig:structure-chi}B how RISM theory
faithfully reproduces the same trends emerging from MC;
in particular, the low-$k$ peak is slightly higher for
$\chi_{\rm s}=0.3$ than for $\chi_{\rm s}=0.4$.
As already discussed in the equimolar case, for the lowest temperatures
RISM underestimates the structural correlations; this can be seen
from the inset of panel B, where the height of the low-$k$ peak is plotted
as a function of concentration.

\section{Conclusions and perspectives}

Using RISM theory and Monte Carlo (MC) simulations,
we have carefully investigated early 
stages of aggregation
in the fluid phase of a colloidal mixture of asymmetric dimers
and spherical particles. All interactions in the model are hard-sphere-like,
except for an additional square-well attraction between the small monomer and the sphere.
Our study is carried out in a temperature regime
where the fluid exhibits local inhomogeneities,
which on cooling evolve into more structured aggregates.
The origin of these spatial modulations can be rationalized in terms of the competition
between a short-range (small monomer-sphere) attraction and a longer-range repulsion
(due to the steric hindrance of large monomers). 
A clear structural indication of the emergence of
spatial inhomogeneities is provided by the development of a low-wavevector peak
in the structure factors of the mixture, as indeed signaled by both theory and simulation. 
In order to acquire microscopic evidence of the existence of aggregates,
we have carried out a cluster analysis of a high number of representative
microscopic configurations taken from simulation.

As a matter of fact, 
as the fluid mixture proceeds towards the full-fledged aggregated state,
the RISM algorithm eventually fails to converge to a physically meaningful solution. 
Despite being only appropriate for moderately inhomogeneous systems, 
the RISM theory agrees reasonably well with MC within its range of applicability, in that it provides reliable
structural indications of the existence of inhomogeneities in the system,
which become fully developed slightly outside
the phase region where the theory is well-behaved.

With the indications acquired from theory, we plan to complement the information obtained
from simulation so as to arrive at a thorough understanding of the mixture behavior
as a function of density, temperature, and concentration. In particular,
it would be interesting to connect the fluid phase properties with 
the low-temperature behavior in the high dilute regime, where 
we have documented a rich phase
scenario with the formation of different supramolecular aggregates, whose
specific nature depends  on the size ratios and
relative concentration of the species.~\cite{Prestipino:19}

A particularly relevant issue when employing
a theory like RISM concerns the possibility to characterize the underlying
microscopic arrangement of an aggregating fluid only in terms of 
structural indicators, namely without explicitly resorting to
microscopic data from simulations. 
Much effort toward this interpretation has been devoted in the field 
of pure SALR fluids, and currently 
three different heuristic structural criteria have been
proposed:  in short, the first 
one is based on the height of
the low-$k$ peak in the static structure factor,
as the fluid progresses within the clustered state;~\cite{Godfrin:13,Godfrin:14}
the second criterion is based on the definition
of a cluster-cluster correlation length,
to be deduced by a Lorentzian fit of the same low-$k$ peak;~\cite{Jadrich:15,Bollinger:16}
the third criterion, involving real-space properties, 
is based on the development of a long-distance shell of neighbors
in the radial distribution function of the fluid~\cite{Bomont:17,Bomont:20a,Bomont:20b}.
It would be highly desirable to verify the possibility to apply 
these criteria also for mixtures. 

Finally, we have introduced a novel method%
~---~based on a microscopic analysis of the high-temperature
regime,
where attraction becomes ineffective~--- to compute the maximum distance
within which two particles can be considered as bonded together.
In forthcoming studies we plan to carefully assess the scope of this
criterion within the realm of cluster-forming fluids.
Should the validity of our approach
not be limited to the case under study,
this would be a noteworthy result since, as also observed in 
the context of SALR particles,
an accurate choice of the bond distance is crucial to properly identify
the self-assembled structures emerging at low temperature.

\section*{Conflicts of interest}
There are no conflicts to declare

%\bibliographystyle{rsc}
%\bibliography{revision}
\providecommand*{\mcitethebibliography}{\thebibliography}
\csname @ifundefined\endcsname{endmcitethebibliography}
{\let\endmcitethebibliography\endthebibliography}{}

\end{document}